# Nonlinear Analog Processing with Anisotropic Nonlinear Films

Michele Cotrufo[1,2,†], Domenico de Ceglia[3,4,†], Hyunseung Jung[5], Igal Brener[5], Dragomir Neshev[6], Costantino De Angelis[3,4], Andrea Alù[1,7]

[1]*Photonics Initiative, Advanced Science Research Center, City University of New York, New York, NY 10031, USA*

[2]*The Institute of Optics, University of Rochester, Rochester, New York 14627, USA*

[3]*CNIT and Department of Information Engineering, University of Brescia, Via Branze, 38, Brescia, 25123, Italy*

[4]*Istituto Nazionale di Ottica, Consiglio Nazionale delle Ricerche, Via Branze, 45, Brescia, 25123, Italy*

[5] *Center for Integrated Nanotechnologies, Sandia National Laboratories, POB 5800, NM 87185, USA*

[6]*ARC Centre of Excellence for Transformative Meta-Optical Systems (TMOS), Research School of Physics, Australian National University*

[7]*Physics Program, Graduate Center of the City University of New York, New York, NY 10016, USA*

[†]*These authors contributed equally*

**Abstract:** Digital signal processing is the cornerstone of several modern-day technologies, yet in multiple applications it faces critical bottlenecks related to memory and speed constraints. Thanks to recent advances in metasurface design and fabrication, light-based analog computing has emerged as a viable option to partially replace or augment digital approaches. Several light-based analog computing functionalities have been demonstrated using patterned flat optical elements, with great opportunities for integration in compact nanophotonic systems. So far, however, the available operations have been restricted to the linear regime, limiting the impact of this technology to a compactification of Fourier optics systems. In this paper, we introduce nonlinear operations to the field of metasurface-based analog optical processing, demonstrating that nonlinear optical phenomena, combined with nonlocality in flat optics, can be leveraged to synthesize kernels beyond linear Fourier optics, paving the way to a broad range of new opportunities. As a practical demonstration, we report the experimental synthesis of a class of nonlinear operations that can be used to realize broadband, polarization-selective analog-domain edge detection.

## Introduction

Flat-optics devices exhibiting a linear and local (LL) response are defined by a spatially varying linear transfer function $T_{LL}(\mathbf{r})$ which depends on the position $\mathbf{r}$, and which can be locally tailored through meta-atoms spatially arranged to form metasurfaces (Fig. 1a). An input signal, defined in real space by a complex electric field amplitude $E_{in}^{\omega}(\mathbf{r})$, is filtered linearly and locally, and transformed into an output field $E_{out}^{\omega}(\mathbf{r}) = T_{LL}(\mathbf{r})E_{in}^{\omega}(\mathbf{r})$. Wavefront engineering is arguably the most prominent application of LL systems:

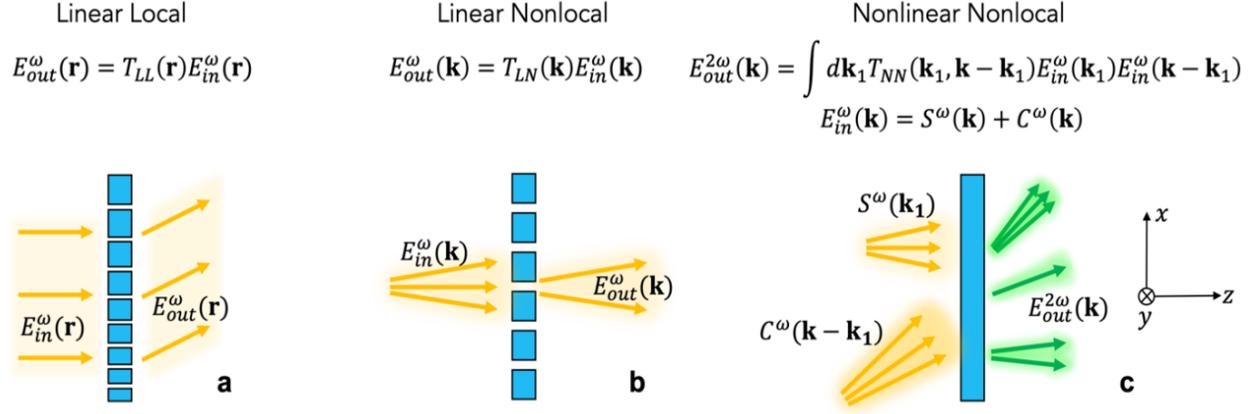

**Fig. 1 Comparison between transfer functions of flat optical elements with LL, LN and NN responses, and typical operations of the three systems. (a)** Beam deflection obtained in a linear and local (LL) system. **(b)** Passive filtering of spatial frequencies obtained in a linear and nonlocal (LN) system. **(c)** A nonlinear and nonlocal (NN) system allows nonlinear mixing of a signal field $S^\omega$ and a control field $C^\omega$, enabling different kinds of nonlinear image processing.

anomalous reflection and refraction for beam steering, focusing and holography have been implemented in a variety of metasurfaces, within plasmonic and dielectric platforms, and in either resonant or non-resonant systems[1]. Spatial dispersion, i.e., nonlocality induced by the long-range coupling of multiple unit cells, is usually treated as a nuisance in LL flat-optics devices, since it complicates their design. However, the nonlocal response of metasurfaces has recently emerged as an additional degree of freedom to achieve advanced functionalities, based on the independent control over spectral and angular selectivity[2]. Nonlocality is usually achieved in patterned flat-optics structures by engineering spatially delocalized resonances, such as guided-mode-resonance gratings and quasi-bound states in the continuum[3]. To date, however, investigations into engineered nonlocalities in metasurfaces have primarily focused on the linear-optics regime. The action of linear nonlocal (LN) systems is well described by a $k$-dependent transfer function $T_{LN}(\mathbf{k})$, which defines how the metasurface filters the angular spectrum $E_{in}^\omega(\mathbf{k})$ of the input function, according to $E_{out}^\omega(\mathbf{k}) = T_{LN}(\mathbf{k})E_{in}^\omega(\mathbf{k})$, as described in Fig. 1b. This transfer function resembles the conventional working principle of Fourier optics systems[4], hence nonlocal metasurfaces have been recently exploited as a platform to compactify linear analog computing and image processing. For example, $n$-th order spatial differentiation[5] can be achieved by tailoring the nonlocality to produce transfer functions $T_{LN}(\mathbf{k}) \propto (i\mathbf{k})^n$, while integration is obtained by synthesizing transfer functions $T_{LN}(\mathbf{k}) \propto (i\mathbf{k})^{1/n}$. LN flat-optics elements readily accommodate image processing functionalities, such as edge detection[6–21] and blurring filters, via high-pass or low-pass spatial-frequency filtering. Despite their appealing features, linear metasurfaces — both local and nonlocal — face limitations related to the restricted numerical aperture and frequency bandwidth[14,22], but most importantly to the inherent linearity of the mathematical operations that can be engineered. In fact, image processing achieved by linear metasurfaces may be also achieved with bulk approaches, such as Fourier 4*f* lens systems. Efforts to overcome these constraints become crucial to push the boundaries of all-optical analog computing metasurfaces and enhancing their capabilities. For instance, deep learning and cryptography inherently require nonlinear functionalities, which cannot be achieved through Fourier optics. Here, we demonstrate that the combination of nonlinear and nonlocal effects in the same flat-optics device offers a powerful strategy to achieve advanced image processing and analog computing functionalities beyond the limits of Fourier systems, offering reduced structural complexity and increased efficiency in terms of angular and frequency bandwidth.

The concept of nonlinear nonlocal (NN) flat-optics is illustrated in Fig. 1c, highlighting the differences with respect to LL and LN approaches. NN flat-optics elements stand out because they can induce nonlinear mixing of spatial and/or temporal frequencies of the input signals. Consider, for example, a scenario in which two fields, a *signal* field $S^\omega(\mathbf{r})$ and a *control* field $C^\omega(\mathbf{r})$, impinge on a NN flat-optics with a large second-order nonlinear response. We denote with $S^\omega(\mathbf{k})$ and $C^\omega(\mathbf{k})$ the angular spectrum decompositions of these fields and, for simplicity, we focus here on the case in which these two fields have the same frequency. The second-harmonic field generated by the total input field, $E_{in}^\omega(\mathbf{k}) = S^\omega(\mathbf{k}) + C^\omega(\mathbf{k})$ can be described by a bi-dimensional transfer function $T_{NN}(\mathbf{k}, \mathbf{k}')$, i.e., $E_{out}^{2\omega}(\mathbf{k}) = \int d\mathbf{k}_1 T_{NN}(\mathbf{k}_1, \mathbf{k} - \mathbf{k}_1) E_{in}^\omega(\mathbf{k}_1) E_{in}^\omega(\mathbf{k} - \mathbf{k}_1)$ (Fig. 1c). Thus, the angular decomposition of the output field, $E_{out}^{2\omega}(\mathbf{k})$, is given by the combination of different contributions involving the bi-dimensional transfer function and the signal and control fields. As a result, the output field contains self-mixing and cross-mixing terms of the input spatial frequencies. By gaining control over the transfer function $T_{NN}(\mathbf{k}, \mathbf{k}')$, nonlinear image processing functionalities can be unlocked. Based on the recent progress in engineering nonlocalities over metasurfaces, it appears that flat optics can form an excellent platform for tailoring the NN transfer functions[23]. Quite remarkably, as we show in the following, even unpatterned thin films can display a nontrivial NN response sufficient to bestow nonlinear image processing capabilities, provided that the nonlinear material of the film supports specific asymmetries[24]. Such functionality can be dynamically controlled through the control signal, unlocking not only nonlinear image processing operations, but also broad reconfigurability, which is not straightforward in linear metasurfaces. In addition, unpatterned thin films featuring nonlinear nonlocality can feature bandwidth and NA much larger than their linear counterparts, offering an exciting platform that can be further augmented by structuring the films in space.

## Results

### Unpatterned flat optics with NN response

To elucidate the potential of combining nonlinearity and nonlocality, we discuss the basic scenario of a nonlinear thin anisotropic film for analog image processing[24]. The geometry in Figure 2 refers to an unpatterned thin film made of a material with second-order nonlinear susceptibility, sufficiently large that we can neglect higher-order nonlinear processes. We consider a two-dimensional problem, with translational invariance of the fields and optical properties along the *y* direction. In addition, we assume that the nonlinear film has a thickness $t$ smaller than the wavelength of the excitation fields ($t \ll \lambda_0$) and it is located at $z = 0$, so that the system can be approximated by a 2D sheet. The nonlinear response of such thin film is mainly dictated by the structure of the second-order nonlinear susceptibility tensor $\chi^{(2)}$, and thus by the chosen material and its crystallographic orientation. Previous works[24] have considered materials for which the only non-zero component of the $\chi^{(2)}$ tensor is $\chi^{(2)}_{zzz}$. Here, we consider a broader and more realistic class of materials, characterized by a nonlinear susceptibility tensor for which only the elements $\chi^{(2)}_{ijk}$ with $i \neq j \neq k$ are non-zero. This is the case of zincblende-type crystals, like gallium arsenide (GaAs) (assuming that the cartesian axis are oriented along the [100], [010] and [001] axis of the crystal), a material platform with large quadratic nonlinearity that has garnered interest in the field of nonlinear photonics for the development of nonlinear metasurfaces for efficient harmonic generation[25–28]. The film is excited by two coherent fields with same frequency: a signal and a control field, both *p*-polarized (electric fields oscillating in the plane *xz*), so that the second harmonic field is *y*-polarized. Hence, second harmonic generation is mediated by the $\chi^{(2)}_{yxz}$ and $\chi^{(2)}_{yzx}$ elements of the nonlinear susceptibility tensor. In this illustrative example, the control field $C^\omega(\mathbf{r})$ takes the form of a Gaussian beam with a large diameter ($\gg \lambda_0$) impinging at oblique incidence with angle $\theta_c$, imparting a linear phase profile across the flat-optics element. The signal field is a small-divergent and normally-incident image modulated in amplitude along the *x* direction by an arbitrary function $s(x)$. For illustrative purposes, we consider a signal made by two adjacent gaussian profiles. The signal and the control fields on the flat-optics element can be written as

$$S^\omega(\mathbf{r}) = [\hat{x}\, s(x)\, S^\omega + \hat{z}\, S_z^\omega]e^{-i\mathbf{k_S}\cdot\mathbf{r}}, \tag{1}$$

$$C^\omega(\mathbf{r}) = \hat{\theta}_c C^\omega\, e^{-i\mathbf{k_C}\cdot\mathbf{r}}, \tag{2}$$

where $\hat{\mathbf{k}}_S = \hat{z}$, $\hat{\mathbf{k}}_C \cdot \hat{x} = \sin\theta_C$ and $\hat{\theta}_c = \cos\theta_C\, \hat{x} - \sin\theta_C\, \hat{z}$. Importantly, the signal field carries a component along the longitudinal direction *z*, associated with the field gradients introduced by the spatial modulation $s(x)$. Indeed, by applying Gauss's law in the absence of charges[29], and adopting the small-divergence approximation ($k_z \approx k_0$), we find $S_z^\omega \approx -\frac{iS^\omega}{k_0}\frac{\partial s(x)}{\partial x}$. Thus, the *z*-component of the signal field is proportional to the first derivative of $s(x)$. The nonlinear film mixes signal and control fields, resulting in a second-harmonic polarization current, in the plane *z=0, given by*

$$\mathbf{P}^{2\omega} = 2\epsilon_0 t \chi^{(2)}_{yxz}[s(x)\, S^\omega + \cos\theta_c\, C^\omega e^{-ik_0 \sin\theta_c x}][S_z^\omega - \sin\theta_c\, C^\omega e^{-ik_0 \sin\theta_c x}]\hat{y}, \tag{3}$$

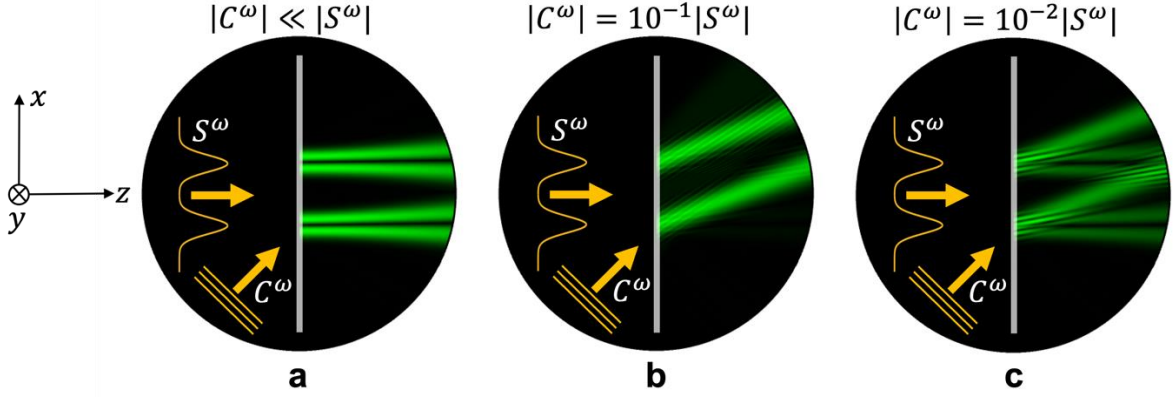

**Fig. 2. Functionalities of a nonlinear nonlocal (NN) flat-optics element made of a thin film of gallium arsenide.** The system is illuminated by two coherent beams: a control (C) and a signal (S), reported in orange. **(a)** When the amplitude of the control is much smaller than the amplitude of the signal, nonlinear edge detection is performed at the SH (the green beam transmitted is the "edge" of the signal, i.e., it is proportional to the derivative of the signal). **(b)** When $|C^\omega| = 0.1|S^\omega|$, nonlinear beam deflection is dominant. **(c)** When $|C^\omega| = 0.01|S^\omega|$, the generated SH signal performs simultaneous nonlinear edge detection and nonlinear beam deflection of the signal. The deflection angle is determined by the angle of incidence of the control signal.

which in turn generates a $y$-polarized second-harmonic field $\mathbf{E}^{2\omega} = G_{yy}\mathbf{P}^{2\omega}$, where $G_{yy} = \frac{2\pi i}{\epsilon_0 \lambda_0}$ is the Green's function of free space. The second-harmonic field contains four terms,

$$\mathbf{E}^{2\omega} = \hat{y}\, 4\pi i \frac{t}{\lambda_0} \chi^{(2)}_{yxz} \big[ s(x) S^\omega S^\omega_z + C^\omega S^\omega_z \cos\theta_c\, e^{-ik_0 \sin\theta_c x} - s(x) S^\omega C^\omega \sin\theta_c\, e^{-ik_0 \sin\theta_c x} - (C^\omega)^2 \sin\theta_c \cos\theta_c\, e^{-i2k_0 \sin\theta_c x} \big], \qquad (4)$$

which correspond to different nonlinear mixing products of the signal and control fields. Starting from this general result, we can now analyze different special scenarios: first, let us consider the case in which the control field is small or absent (i.e., $|C^\omega| \ll |S^\omega|$), as in Fig. 2a. The first term in Eq. (4), proportional to $s(x)\, S^\omega S^\omega_z$, is dominant in this case, and it is the only one that emerges perpendicularly with respect to the film. The corresponding SHG field, emerging from the right side (green pattern), contains four peaks, corresponding to the points where the two input Gaussian peaks have the largest slopes. Since the SHG field amplitude is proportional to the signal gradients, this operation can be used to realize a nonlinear form of edge detection on the input function $s(x)$.

We emphasize that the edge detection obtained here (Fig. 2a) is not only based on a different mathematical operation, nonlinear in nature, than in the case of Fourier optical elements, but it is also based on a fundamentally different phenomenon than the one obtained in linear metasurfaces.[6–21] Here, edge detection is obtained by exploiting the anisotropy of the $\chi^{(2)}$ tensor of GaAs and the corresponding polarization- and angle-selective response with respect to free-space radiations. In the case of GaAs, the SH field is proportional to the longitudinal component of the input field, which in turn carries information about the spatial derivative of the image. This is in contrast to linear approaches to edge detection, which rely on Fourier filtering performed by linear metasurfaces[3,5,12,14,19]. In those systems, a dispersive resonance is engineered in order to induce a strong variation of the linear transmission with angle of incidence, which then leads to a Laplacian-type operator in the Fourier domain. Because of this mechanism and the need for high-Q resonances, these linear devices are inherently narrowband. In contrast, the output SH field generated by the NN response of the thin GaAs film contains the product of the signal by its first derivative – a nonlinear operation, which cannot be obtained with any linear system. In addition, the approach demonstrated here (Eqs. 1-4) does not require a resonance or tailored dispersion, and thus it inherently benefits from a broad operational spectral bandwidth. The bandwidth is only limited by the possible onset of absorption inside the thin layer for photons with energy above the material bandgap, as we report experimentally in the next section.

The second and third terms in Eq. (4) are cross-terms produced by the spatial-frequency mixing of control and signal fields. They form a beam that propagates in the direction $\theta_{CS} = \mathrm{asin}\left(\frac{\sin\theta_C}{2}\right)$. In the small divergence approximation ($k_z \approx k_0 \gg \left|\frac{\partial}{\partial x}\right|$), the main contribution contained in this beam is the third term in Eq. (4), proportional to $-s(x)\, S^\omega C^\omega \sin\theta_c\, e^{-ik_0 \sin\theta_c x}$. This term is a scaled version of the input signal deflected towards $\theta_{CS}$. Importantly, the deflection angle $\theta_{CS}$ is solely determined by the incident angle of the control signal, whereas the amplitude of the control signal may be tuned to make the beam-deflection effect dominant. For example, in Fig. 2b a nonlinear beam deflection can be clearly seen for $|C^\omega| = 0.1|S^\omega|$.

Another interesting scenario is displayed in Fig. 2c for the case $|C^\omega| = 0.01|S^\omega|$. Here, the nonlinear edge-detection in the normal direction and the deflection of the input image at the angle $\theta_{CS}$ happen simultaneously. Such a regime may be of interest for applications where simultaneous access to the unprocessed and processed images is required. The fourth term in Eq. 4 is transversely phase-matched to the control plane wave, and therefore, it is the only one that emerges at the angle $\theta_C$. This term is not particularly interesting, and it becomes dominant only when $|C^\omega| \gg |S^\omega|$.

The examples discussed in Fig. 2 showcase various intriguing capabilities of NN systems, even in the most basic scenario of an unpatterned thin film: (i) image processing functionalities that typically require patterned metasurfaces in the linear regime, such as beam deflection (or anomalous refraction) and edge detection, may be achieved by leveraging the nonlinear and nonlocal response of unpatterned thin films; (ii) the response of an NN system is intensity and control-beam dependent, hence tunable – in the example above one can switch the functionality from beam defection to edge detection by varying the ratio of signal and control amplitudes; (iii) the resulting operations are inherently nonlinear – for instance, edge detection here is nonlinear and fundamentally different from the Laplacian produced by a linear optical system, as it results from the multiplication of two fields, specifically, the signal itself multiplied by its derivative; (iv) since this approach does not rely on a resonance, it is also free from fundamental constraints on the operational spectral bandwidth, as it happens instead for linear devices based on resonances or tailored dispersion; (v) more complex nonlinear functionalities may be realized by leveraging the control function as a means to write functionalities – while in the reported example, the control is only used to achieve nonlinear beam deflection, for more sophisticated functionalities, such as nonlinear holography, the control may be a structured beam with arbitrary amplitude and phase profile; (vi) by locally structuring the metasurface we may induce engineered resonances that trade bandwidth for efficiency, and we may engineer the nonlocality of the $\chi^{(2)}$ tensor, opening the design space to more complex nonlinear operators.

**Edge detection in nonlinear unpatterned films**

Edge detection with NN flat-optics has been theoretically proposed[24] based on a thin layer of nonlinear material with susceptibility tensor featuring only the $\chi^{(2)}_{zzz}$ element. This type of response can be achieved, for example, at interfaces[30], in nanolaminate films[31], and in quantum well systems[32]. Here, based on the general formalism described above, we demonstrate theoretically and experimentally that nonlinear edge detection is not limited to $\chi^{(2)}_{zzz}$. By contrast, we show that broadband edge detection arises in unpatterned flat-optics with any type of nonlinear susceptibility tensor that involves at least one longitudinal component of the input pump field. Building upon the example outlined in the previous section, we consider a NN system comprising an unpatterned thin film of (001) GaAs (Fig. 3a). For nonlinear edge detection (Fig. 2a), the control signal is not necessary. To understand the mechanism of nonlinear edge detection in the GaAs flat-optics system, we assume that the input image, carried by a wave at the fundamental frequency (FF), is projected on the nonlinear film. Using the same approach outlined in the previous section, the FF pump can be written as: $\mathbf{E}^\omega = \mathbf{E}^\omega_\perp + E^\omega_z \hat{z}$, where the longitudinal component is $E^\omega_z \approx \frac{i}{k}\left(\frac{\partial E^\omega_x}{\partial x} + \frac{\partial E^\omega_y}{\partial y}\right)$ [29]. It follows that any nonlinear flat-optics system that is sensitive to $E^\omega_z$ — i.e., which can generate a nonlinear polarization when a $z$-polarized pump field is present — will produce a nonlinear signal proportional to the spatial derivative of the pump field. One of these systems is the unpatterned GaAs film considered here. Indeed, owing to its zincblende crystal structure, the nonlinear susceptibility tensor of GaAs obeys $t^{-1}\chi^{(2)} \equiv \chi^{(2)}_{xyz} = \chi^{(2)}_{yxz} = \chi^{(2)}_{zxy} = \chi^{(2)}_{xzy} = \chi^{(2)}_{yxz} = \chi^{(2)}_{zyz} \neq 0$, while all the other components vanish [31].

Using the Green's function approach[24,33], we can retrieve the relationship between the transverse part of the second-harmonic (SH) field and the transverse part of the FF field. The SH field, evaluated at the sheet position, reads

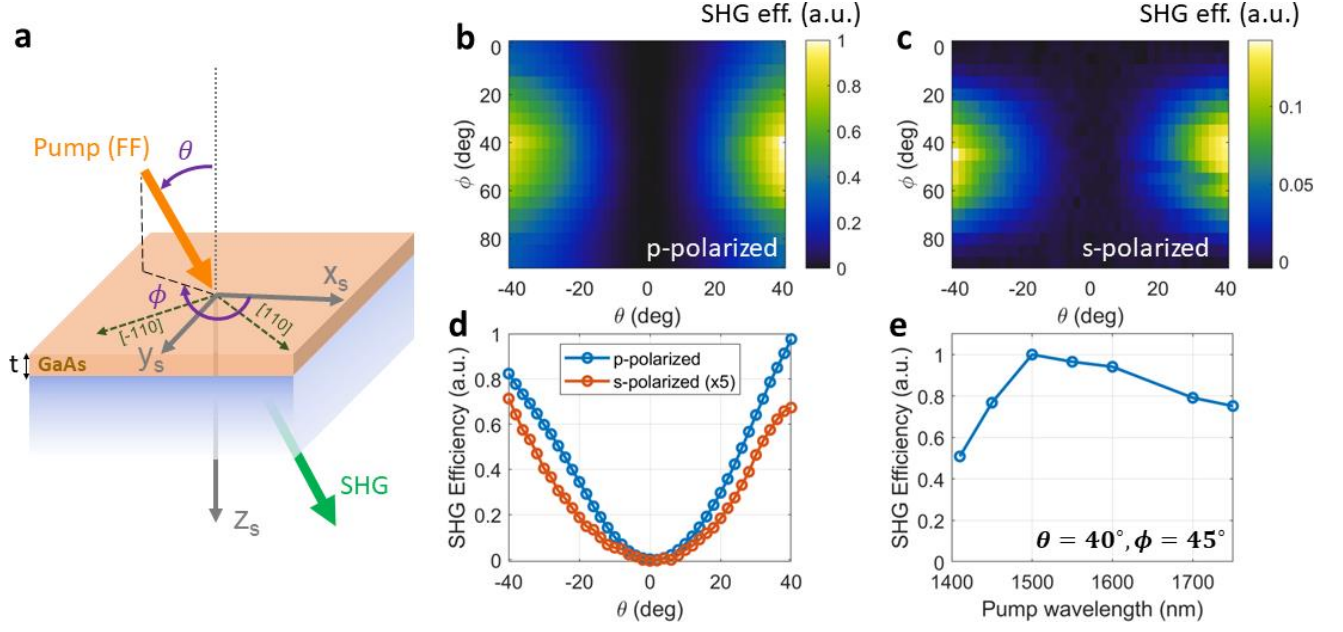

**Figure 3. Sample and preliminary characterization. (a)** Schematic of the sample and measurement. A thin layer of GaAs is bonded on a quartz substrate. The axes $x_s$, $y_s$, and $z_s$ are aligned to the [100], [010] and [001] crystal directions of the sample, respectively. The sample is excited by a pump beam (orange arrow) at the fundamental frequency (FF) impinging along a direction defined by the polar ($\theta$) and azimuthal ($\phi$) angle, and with either s or p polarization. The second-harmonic emission (green arrow) generated along a direction parallel to the pump is collected on the opposite side of the sample. **(b-c)** Measured SHG efficiency (arbitrary units) for pump wavelength $\lambda_{FF} = 1550$ nm, versus $\theta$ and $\phi$, for a *p*-polarized (panel b) and *s*-polarized (panel c) pump. **(d)** Measured SHG efficiency versus $\theta$, for pump wavelength $\lambda_{FF} = 1550$ nm and $\phi = 45°$. **(e)** Measured SHG efficiency versus pump wavelength for $\theta = 40°$, $\phi = 45°$ and p-polarized pump.

$$E_x^{2\omega} = -\chi^{(2)} \left[ 2 E_y^\omega \frac{\partial E_x^\omega}{\partial x} + E_y^\omega \frac{\partial E_y^\omega}{\partial y} + E_x^\omega \frac{\partial E_y^\omega}{\partial x} \right] \quad (5)$$

$$E_y^{2\omega} = -\chi^{(2)} \left[ 2 E_x^\omega \frac{\partial E_y^\omega}{\partial y} + E_x^\omega \frac{\partial E_x^\omega}{\partial x} + E_y^\omega \frac{\partial E_x^\omega}{\partial y} \right]. \quad (6)$$

The expressions in Eq. (5) and Eq. (6), derived in the Methods, unveil the nonlinear edge detection operation of the NN film. All the terms in the SH fields are proportional to the first-order spatial derivatives of the pump signal multiplied by a component of the signal itself, confirming the nonlinear high-pass filtering operation illustrated in the previous section. An interesting feature of the nonlinear edge detection operation is the role of the polarization state of the pump, which may be used as a control knob to highlight selectively edges with different orientation. To show this, let us first assume that the signal is linearly polarized along *x*, so that the transverse FF signal is $\mathbf{E}_\perp^\omega = E_x^\omega(x,y)\hat{x}$. Since $E_y^\omega = 0$, from Eq. (5) and (6) it follows that the resulting SH field is *y*-polarized and equal to $E_y^{2\omega} = -\chi^{(2)} E_x^\omega \frac{\partial E_x^\omega}{\partial x}$. Therefore, large SH signals will be generated by features of the input image with non-zero values of the spatial derivative $\frac{\partial E_x^\omega}{\partial x}$; such scenario is obtained, for example, when the input image contains edges that are parallel to *y*. Instead, edges that are parallel to *x* will lead to nonzero values of only the derivative $\frac{\partial E_x^\omega}{\partial y}$, which in this case does not contribute to the SHG efficiency. Conversely, when the pump signal is *y*-polarized, the SH field is *x*-polarized and equal to $E_x^{2\omega} = -\chi^{(2)} E_y^\omega \frac{\partial E_y^\omega}{\partial y}$. In this scenario, only horizontal edges (parallel to *x*) will give rise to high SH signals. For circular polarization $\mathbf{E}_\perp^\omega = E_0(x,y)(\hat{x} \pm i\,\hat{y})/\sqrt{2}$, with the plus (minus) sign indicating right (left) handedness, the SH field is

$$E_x^{2\omega} = -\frac{1}{\sqrt{2}}\chi^{(2)}\left[\pm 3iE_0 \frac{\partial E_0}{\partial x} - E_0 \frac{\partial E_0}{\partial y}\right] \qquad (7)$$

$$E_y^{2\omega} = -\frac{1}{\sqrt{2}}\chi^{(2)}\left[\pm 3iE_0 \frac{\partial E_0}{\partial y} + E_0 \frac{\partial E_0}{\partial x}\right]. \qquad (8)$$

Therefore, all edges of the FF image are uniformly detected in the generated output image. It is important to emphasize once again the relevant differences between the results obtained here and the edge detection obtained in linear metasurfaces by engineering dispersive optical modes. First, the nonlinear spatial differentiation in Eqs. 7-8 (i.e., under circularly polarized pump) is isotropic with respect to azimuthal rotations of the input image – i.e., if the input image is rotated within the *xy* plane, the output image will be rotated as well, without any image distortion (see also Supplementary Information, Section S4). By contrast, edge detection obtained via Fourier filtering in linear metasurfaces often suffers from poor isotropy[10,12], due to the fact that a metasurface made of periodic patterns cannot be invariant under arbitrary rotations. The most isotropic responses are obtained with metasurfaces with $C_6$ rotational symmetry,[14] within moderately large ranges of impinging polar angles. Yet, high anisotropies are inevitable for larger impinging angles[14]. Second, the output fields in Eqs. 5-8 are proportional to the first-order derivative of the input field. First-order (or, in general, odd-order) differentiation is challenging to achieve in linear patterned metasurfaces, because it requires a unit cell that simultaneous breaks vertical and horizontal symmetry[5], introducing significant fabrication challenges. Instead, the proposed NN thin film provides first-order differentiation naturally, without requiring any symmetry breaking and challenging fabrication processes.

We experimentally demonstrate this nonlinear image processing operation based on the sample shown in Fig. 3a, consisting of a 480-nm-thick flat slab of (001) GaAs bonded on a thick quartz substrate. To verify the capability of this thin, unpatterned slab to provide the desired nonlinear image processing operation, we first measured how the SHG efficiency depends on the pump direction, polarization and wavelength. The sample was illuminated with a quasi-plane-wave pump excitation at the fundamental frequency (wavelength $\lambda_{FF} \in [1400, 1750]$ nm), provided by a tunable pulsed laser (pulse duration 2 ps, repetition rate 80 MHz). The generated SH signal was collected along the same direction as the pump (see Methods for details on the setup). No polarization selection was performed on the collected SH signal. In Figs. 3(b-c), we show the measured SHG efficiency as a function of the polar ($\theta$) and azimuthal ($\phi$) angles of the pump direction (see Fig. 3a for the definition of the angles), for a fixed pump wavelength $\lambda_{FF} = 1550$ nm and for *p*-polarization (Fig. 3b) and *s*-polarization (Fig. 3c) (see Methods for additional details on the measurement protocol). The measured efficiency follows the expected patterns for both polarizations (see Supplementary Information, Section S3, for a comparison between these experiments and numerically calculated data). Figure 3d shows a sliced 1D data for $\phi = 45°$ for both polarizations. The measurements confirm that pumps propagating at normal incidence ($\theta = 0°$) — which in the imaging experiment carry the information on the DC components of the input image — do not lead to any SH generation. As the excitation polar angle $\theta$ increases, the SH signal increases monotonically. For *s* polarization, no SH emission is observed when $\phi = n\pi/2$, as expected from the symmetry of the $\chi^{(2)}$ tensor. For *p*-polarization, instead, the SH emission is nonzero for any azimuthal angle, albeit with some large modulation.

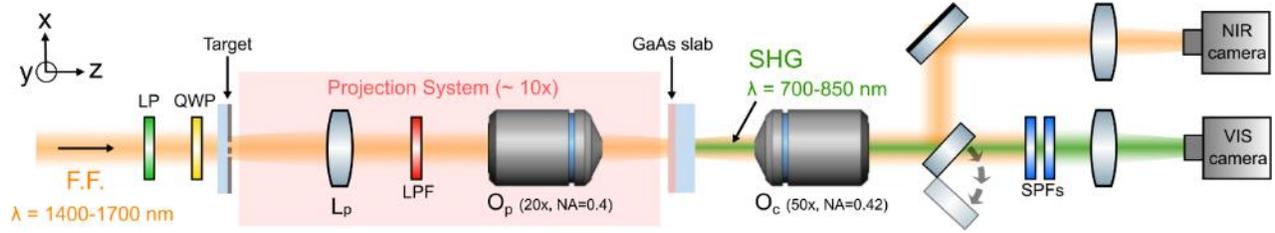

**Figure 4. Setup for nonlinear image processing measurements.** A pulsed laser with wavelength $\lambda_{FF}$ is prepared in a certain polarization state and then used to illuminate a mask ('Target') to create an optical image. A projection system formed by a lens ($L_p$) and an objective ($O_p$, 20x, NA=0.4) relays the image on the GaAs slab. A long pass filter (LPF) inside the projection system rejects any SH signal generated by the target material. The signal propagating on the other side of the GaAs slab (which contains both the FF and the SHG image) is collected by a second objective ($O_c$, 50x, NA = 0.42) and redirected either to a near-infrared camera or to a visible camera. LP = Linear Polarizer, QWP = Quarter Wave Plate, LPF = Long Pass Filter, SPFs = Short Pass Filters.

Importantly, the SHG efficiency depends weakly on the pump wavelength, as shown in Fig. 3e. Here we measured the SHG efficiency versus pump wavelength for fixed $\theta = 40°$, $\phi = 45°$ and *p*-polarized pump. The SHG efficiency remains above 50% of its peak value for any $\lambda_{FF} \in [1400, 1750]$ nm. The weak modulation of SHG efficiency as a function of the pump wavelength is due to a combination of the frequency dependence of the $\chi^{(2)}$ of GaAs[34,35] and the weak Fabry-Perot resonances created by the finite slab, as confirmed by numerical simulations (see Supplementary Information). The decrease in efficiency for $\lambda_{FF} < 1450$ nm is attributed to the onset of absorption of the corresponding SH signal in GaAs. In the Supplementary Information (Section S3), we show numerically calculated data corresponding to the measurements in Fig. 3(b-d), which display excellent agreement with the experiments.

After having verified that our sample can provide the expected nonlinear transfer function, we tested its behavior as a flat optics for nonlinear image processing by using the setup schematized in Fig. 4, and discussed in more detail in the Methods. The pump signal was provided by the same tunable pulsed laser used for the measurements in Fig. 3. The power and polarization states of the pump were controlled by a series of polarization optics. The collimated laser was then used to illuminate a mask (labelled 'target' in Fig. 4), which contains different test images. The typical average power on the target was of the order of 1 W, corresponding to a pulse energy of the order of ~10 nJ. The image created by the target was relayed onto the GaAs slab with a projection system (red-shaded box) formed by the lens $L_p$ and the objective $O_p$ (20x, NA=0.4). The projection system shrinks the image size by a factor of approximately $10 \times$, thus increasing the local pump intensity. A long pass filter (LPF) inside the projection system rejected any SH signal generated by the material forming the mask. The image at the fundamental frequency impinged on the GaAs slab from one side, and the signal emerging from the other side of the slab (containing both the FF and the SH images) was collected by another objective ($O_c$, 50x, NA=0.42). This signal was then redirected either to a NIR camera to record the FF image or to a visible camera to record the SH image. In the second case, the strong FF signal was filtered out by two short-pass filters (SPFs). In this experiment, the GaAs slab was oriented such that the [110] direction of GaAs was parallel to the y-direction of the lab reference frame (defined in Fig. 4), while the [-110] direction of GaAs was parallel to the x-direction defined in Fig. 4).

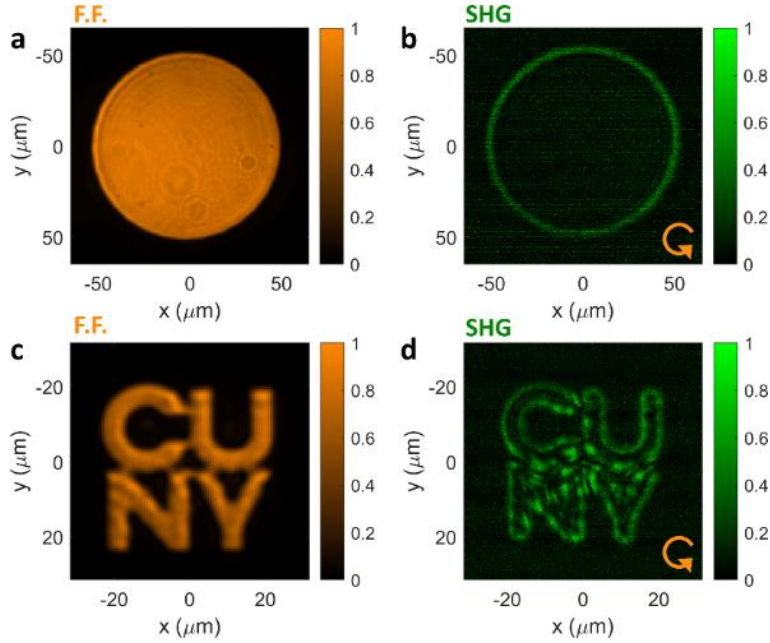

**Figure 5. Experimental nonlinear image processing.** Nonlinear image processing of different input images carried by a circularly polarized pump with $\lambda_{FF} = 1550$ nm. **(a-b)** The input image (panel a) is a circle with a diameter of approximately 100 μm. The image recorded at the SH wavelength (panel b) displays the expected signal processing, i.e. an isotropic enhancement of the edges. **(c-d)** Same as in panels a-b, but with an input image made of the logo of one of our institutions.

In Fig. 5, we show the experimental results with input images at a pump wavelength $\lambda_{FF} = 1550$ nm. In this figure, the pump electric field is circularly polarized, which, based on Eqs. 7-8, ensures the enhancement of all edges, independently of their orientation. Here and in all following results, the spatial coordinates on the axis of all figures correspond to calibrated spatial dimensions measured in the plane of the GaAs slab. In Fig. 5(a-b), the input image (Fig. 5a) is a circle with a diameter of 100 μm. The corresponding SHG image (Fig. 5b) shows the expected edge-detection functionality: the SH signal is strongly enhanced along the edges of the input images, while it is almost zero in the spatial regions where the intensity of the input image is constant. Similar results are obtained with a different input image, consisting of the logo of one of our institutions (Fig. 5(c-d)).

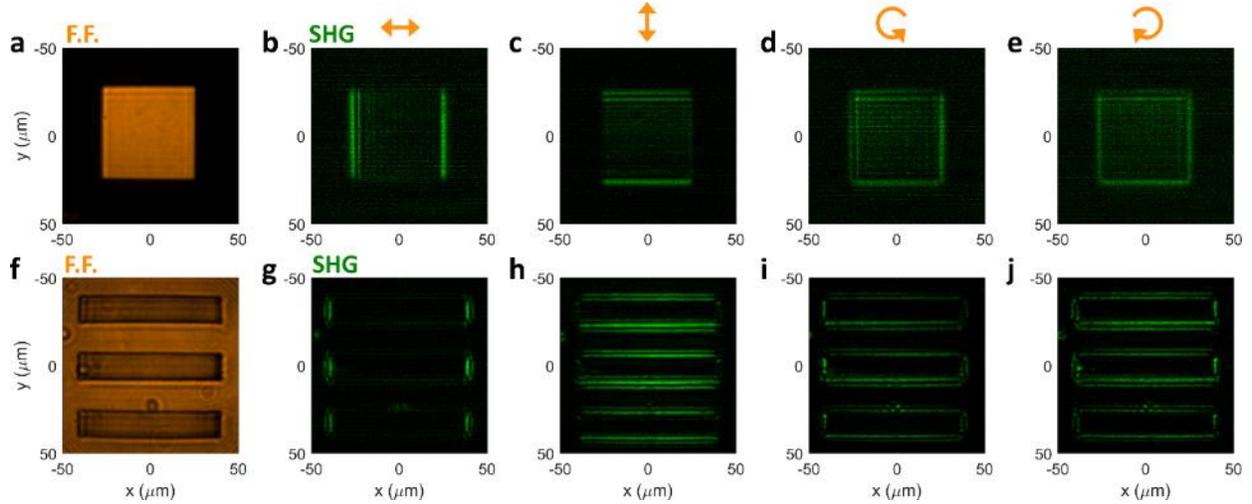

**Figure 6. Dependence on input polarization of nonlinear image processing. (a-e)** An input image (panel a) containing a 'positive' square (i.e., high intensity inside the square) is processed by the GaAs slab. We recorded the generated SH image for different polarization of the input image: horizontal (panel b), vertical (panel c), right circularly polarized (panel d), and left circularly polarized (panel e). **(f-j)** Same as in panels (a-e), but for the case in which the input image consists of three 'negative' rectangles (i.e. zero intensity inside the rectangles).

Next, we investigate the dependence of the features of the output image on the polarization of the input image. To this aim, we consider input images with well-defined edge orientations, such as squares and rectangles. In Fig. 6(a-e), we consider a fixed input FF image formed by a square with a width of about 50 μm (Fig. 6a). We recorded the generated SH image for different input polarizations: horizontal (Fig. 6b), vertical (Fig. 6c), right circularly polarized (Fig. 6d), and left circularly polarized (Fig. 6e). As expected from the discussion above, when the input signal is linearly polarized (Figs. 6b and 6c) the edges of the input image which are oriented orthogonal to the input polarization are maximally enhanced in the output images, while edges that are parallel to the input polarization are absent in the output images. Instead, when using circular polarization (with either chirality), all edges are uniformly enhanced (Figs. 6d-e). The results in Fig. 6(a-e) were obtained with a "positive" image, i.e., a shape whose internal area has a large intensity, while the external area has zero intensity. We have also verified that the same image processing and polarization-dependent behavior is obtained with "negative" images, i.e. images composed of dark shapes surrounded by a high-intensity background (Figs. 6(f-j)).

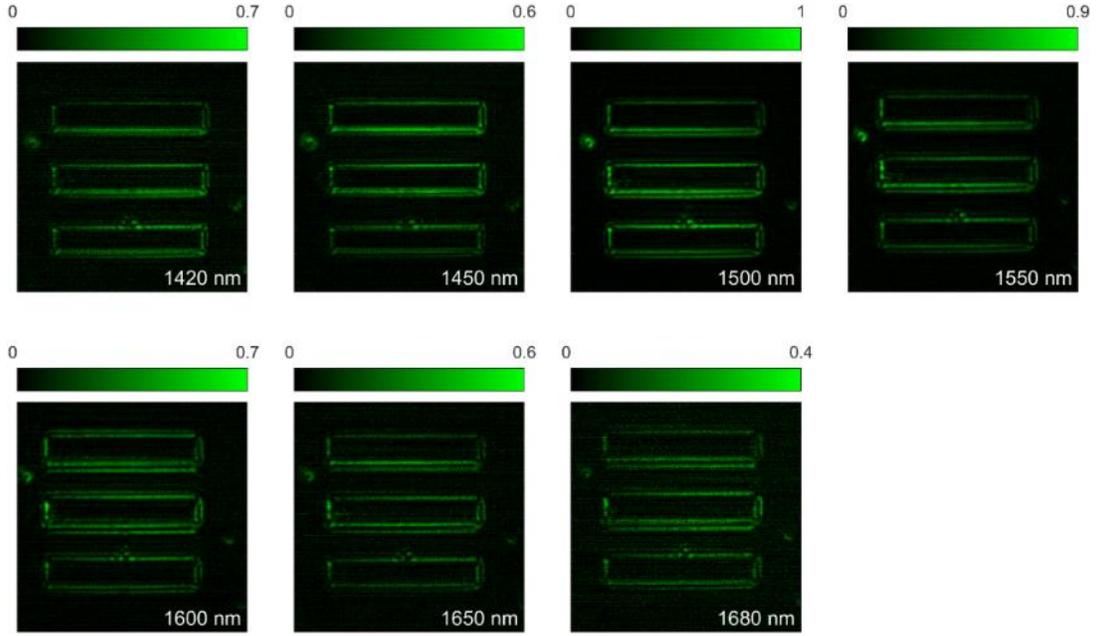

**Figure 7. Nonlinear image processing versus input wavelength.** The input image is the same as in Fig. 5f, and the input polarization is circular. The seven panels in the figure show the recorded SHG image as the input wavelength was varied between 1420 nm and 1680 nm. The input wavelength is denoted in the left-bottom corner of each panel. In each measurement, the intensity recorded by the camera was normalized by the squared of the input power. Then, all data was further normalized with the respect to the maximum peak intensity, which occurs for $\lambda_{FF} = 1500$ nm.

Finally, we show that our sample can perform nonlinear image processing over an extremely broad range of input frequencies. In Fig. 7, we used the same input image as in Figs. 6(f-j), and we fixed the input polarization to circular. We varied the input wavelength from $\lambda_{FF} = 1420$ nm to $\lambda_{FF} = 1680$ nm, and recorded the corresponding SH output images. As shown by the different panels in Fig. 7, all output images display well-defined edge detection. The output image is essentially independent of the FF wavelength (indicated in the bottom-right corner of each panel of Fig. 7). To quantitatively compare the output intensities and the edge detection efficiency at different input wavelengths, all SHG images in Fig. 7 were acquired in the same experimental conditions, and the pixel intensities of each measurement have been normalized by the square of the corresponding FF power. Furthermore, to aid the comparison between different measurements, we normalized all data with respect to the highest peak intensity among all panels in Fig. 7. With this procedure, the intensities of the different output images in Fig. 7 can be readily compared by looking at the upper bound of the corresponding colorbar. As expected from the efficiency curve in Fig. 3e, the highest peak intensity in Fig. 7 occurs for $\lambda_{FF} = 1500$ nm. Moreover, the variation of the peak intensity with input wavelength is fairly weak, and the peak intensity remains above 0.7 for $\lambda_{FF} = 1420$ nm and above 0.4 for $\lambda_{FF} = 1680$ nm. We note that the wavelength range considered in Fig. 7 (1420 nm – 1680 nm) was solely determined by the power available in our laser system, and we expect the operational bandwidth to be even larger, ultimately limited only by the increased absorption of the SH within the GaAs slab as the SH wavelength $\lambda_{FF}/2$ further decreases. In the Supplementary Information (Sections S1-S3), we discuss additional numerical calculations of the SHG efficiency of our GaAs slab versus pump wavelength.

## Discussion and Outlook

In this work, we have proposed and experimentally demonstrated that, by combining nonlinearities with nonlocalities in unpatterned thin films, it is possible to perform analog nonlinear image processing tasks, unlocking fundamental operations and practical metrics of interest that are not available in linear patterned metasurfaces. Based on these principles, we have provided an experimental demonstration of broadband, high-resolution nonlinear edge detection with a thin slab of Gallium Arsenide.

From a fundamental standpoint, our approach allows performing nonlinear image processing operations – such as the product of the input image by its own derivative. Such operation cannot be implemented in linear devices based on Fourier optics, and can provide an effective scheme for phase-imaging based on which rapid phase changes are emphasized on top of an existing image. Remarkably, this operation is obtained within a thin unpatterned film, without the need of any nanostructuring, in strong contrast to linear devices which require nanoscale patterning to induce the required *k*-space transfer function. We instead leverage the inherent nonlocality of an anisotropic thin film. Besides the fundamental difference in working principle and achievable computational tasks, our approach also boosts several key-metrics compared to Fourier-based linear metasurface approaches. Indeed, since our approach does not rely on optical resonances, it is inherently broadband. In our experiment, we demonstrated nonlinear edge detection over a 260 nm bandwidth, only limited by the practical constraints of our setup. Besides the increased spectral bandwidth, the absence of any resonant effect also has other important benefits. First, a non-resonant optical response lifts any strong constraint on the numerical aperture of our device, overcoming typical challenges which instead affect patterned linear metasurfaces that rely on dispersive resonances[14]. Second, the absence of resonant effects largely benefits the power handling capability of our flat device as compared to resonant nonlinear metasurfaces. Indeed, metasurfaces made out of sub-wavelength resonators typically lead to a resonantly enhanced electric field inside the resonators, which increases the thermal damage induced by free-carrier injection and thus strongly lowers the damage threshold[27]. Instead, our device is completely homogeneous within the in-plane dimensions, thus avoiding any local enhancement of the electric field intensity. This feature dramatically increases the power damage threshold, making it comparable to the one of a bulk crystal. While we trade these advantages for power efficiency – as nonlinear optical processes in thin films are very weak – the SH fields are dominated by the processed image, hence they can be easily resolved even without strong output power levels.

More broadly, the phenomena demonstrated here pave the way for the use of flat-optics for nonlinear analog computation, with potential applications in neuromorphic computing, optics-based neural networks, and nonlinear signal processing. In this work, we focused on a flat unpatterned layer, where the nonlinear transfer function $T_{NN}(\mathbf{k}, \mathbf{k}')$ is solely dictated by the anisotropy of the $\chi^{(2)}$ tensor. We envision that this approach can be extended to nonlinear patterned metasurfaces, whereby the nonlinear transfer function can be further tailored by optical resonances to achieve more advanced nonlinear optical functionalities.

## Methods

### Second-harmonic generation from a thin sheet of GaAs

The nonlinear thin film is modeled as a flat-optics element, or sheet, with quadratic nonlinearity. The sheet is located at $z = 0$. Second-harmonic light emission from the sheet is due to of the interaction of its nonlinear susceptibility with the fundamental-frequency field (or pump field) at $z = 0$. Let's analyze the quadratic nonlinear response associated with an arbitrary $\chi^{(2)}_{\ell m n}$ element, where $\ell$, $m$ and $n$ indicate Cartesian coordinates. In the Fourier-space representation, the **k**-spectrum of the input pump field (at frequency $\omega$) in the plane of the sheet ($z = 0$) is expressed as a two-dimensional Fourier transform, $\tilde{E}^{\omega}_i(\mathbf{k}_\perp, 0) = \mathfrak{I}[E^{\omega}_i(\mathbf{r}_\perp, 0)]$, where $i = x, y, z$, $\mathbf{r}_\perp = (x, y)$ is the direct space in-plane coordinate and $\mathbf{k}_\perp = (k_x, k_y)$ is the spatial-frequency in-plane vector. If the nonlinear susceptibility $\chi^{(2)}_{\ell m n}$ does not introduce angular dispersion, the **k**-spectrum of the second-harmonic field (at frequency $2\omega$) at any longitudinal position $z$ can be calculated as a convolution of the pump signal field components[24,33], which in Cartesian coordinates is written as

$$\tilde{E}^{2\omega}_i(\mathbf{k}_\perp, z^\pm) = \tilde{G}_{i\ell}(\mathbf{k}_\perp, z^\pm) \chi^{(2)}_{\ell m n} \tilde{E}^{\omega}_m(\mathbf{k}_\perp, 0) * \tilde{E}^{\omega}_n(\mathbf{k}_\perp, 0) \tag{M1}$$

where the star symbol is the convolution operator and

$$\tilde{\mathbf{G}}(\mathbf{k}_\perp, z^\pm) = \frac{ie^{\pm ik_z z}}{2k_z} \begin{pmatrix} \frac{k_y^2 k^2 + k_x^2 k_z^2}{k_x^2 + k_y^2} & \frac{k_x k_y k_z^2 - k_x k_y k^2}{k_x^2 + k_y^2} & \mp k_x k_z \\ \frac{k_x k_y k_z^2 - k_x k_y k^2}{k_x^2 + k_y^2} & \frac{k_x^2 k^2 + k_y^2 k_z^2}{k_x^2 + k_y^2} & \mp k_y k_z \\ \mp k_x k_z & \mp k_y k_z & k_x^2 + k_y^2 \end{pmatrix} \tag{M2}$$

indicates the spectral Green's function. Here, $k_z = \sqrt{k^2 - k_x^2 - k_y^2}$ is the longitudinal component of the wavevector and $k$ is the wavenumber. If the pump signal has a small spatial-frequency spectrum in the transverse plane (corresponding to an input image with features much larger than the free-space wavelength), then $k_z \approx k$ and the expression in Eq. (M1) simplifies to

$$\tilde{E}_i^{2\omega}(\mathbf{k}_\perp, z^\pm) = \frac{ie^{\pm jkz}}{2k} \begin{pmatrix} k^2 & 0 & \mp k_x k \\ 0 & k^2 & \mp k_y k \\ \mp k_x k & \mp k_y k & k_x^2 + k_y^2 \end{pmatrix} \chi_{\ell mn}^{(2)} \tilde{E}_m^\omega(\mathbf{k}_\perp, 0) * \tilde{E}_n^\omega(\mathbf{k}_\perp, 0). \tag{M3}$$

In the case of GaAs, the only nonzero $\chi_{\ell mn}^{(2)}$ elements are those for which $\ell \neq m \neq n$, namely $t^{-1}\chi^{(2)} \equiv \chi_{xyz}^{(2)} = \chi_{yxz}^{(2)} = \chi_{zxy}^{(2)} = \chi_{xzy}^{(2)} = \chi_{yxz}^{(2)} = \chi_{zyz}^{(2)}$, where $t \ll \lambda_0$ is the thickness of the flat-optics element. Therefore, the transverse components of the transmitted second-harmonic field can be written in Fourier space as:

$$\tilde{E}_x^{2\omega}(\mathbf{k}_\perp, z^\pm) = i\chi^{(2)} k e^{jkz} \left( \tilde{E}_y^\omega(\mathbf{k}_\perp, 0) * \tilde{E}_z^\omega(\mathbf{k}_\perp, 0) - \frac{k_x}{k} \tilde{E}_x^\omega(\mathbf{k}_\perp, 0) * \tilde{E}_y^\omega(\mathbf{k}_\perp, 0) \right) \tag{M4}$$

$$\tilde{E}_y^{2\omega}(\mathbf{k}_\perp, z^\pm) = i\chi^{(2)} k e^{jkz} \left( \tilde{E}_x^\omega(\mathbf{k}_\perp, 0) * \tilde{E}_z^\omega(\mathbf{k}_\perp, 0) - \frac{k_y}{k} \tilde{E}_x^\omega(\mathbf{k}_\perp, 0) * \tilde{E}_y^\omega(\mathbf{k}_\perp, 0) \right) \tag{M5}$$

The transverse components of the second-harmonic field at $z = 0^+$, i.e., in transmission, can be retrieved by anti-Fourier transforming the expressions in Eq. (M4) and (M5):

$$E_x^{2\omega}(\mathbf{r}_\perp, 0^+) = i\chi^{(2)} k E_y^\omega(\mathbf{r}_\perp, 0) E_z^\omega(\mathbf{r}_\perp, 0) - \chi^{(2)} \frac{\partial \left( E_x^\omega(\mathbf{r}_\perp, 0) E_y^\omega(\mathbf{r}_\perp, 0) \right)}{\partial x} \tag{M6}$$

$$E_y^{2\omega}(\mathbf{r}_\perp, 0^+) = i\chi^{(2)} k E_x^\omega(\mathbf{r}_\perp, 0) E_z^\omega(\mathbf{r}_\perp, 0) - \chi^{(2)} \frac{\partial \left( E_x^\omega(\mathbf{r}_\perp, 0) E_y^\omega(\mathbf{r}_\perp, 0) \right)}{\partial y} \tag{M7}$$

where the anti-transformations $ik_x \to \partial/\partial x$ and $ik_y \to \partial/\partial y$ have been applied. Introducing the expression of the longitudinal field as a function of the derivative of the transverse field, $E_z^\omega \approx \frac{i}{k}\left(\frac{\partial E_x^\omega}{\partial x} + \frac{\partial E_y^\omega}{\partial y}\right)$, in Eq. (M6) and Eq. (M7) leads to the expressions reported in Eq. (5), Eq. (6), Eq. (7) and Eq. (8).

**Sample Fabrication**

A 480-nm-thick GaAs layer is epitaxially grown on a (001)-oriented GaAs wafer with stop-etch layers. The stop-etch layers consist of 120-nm-thick Al0.55Ga0.45As, 100-nm-thick GaAs, and 20-nm-thick Al0.55Ga0.45As layers, in that order. The wafer is diced to be 10 x 10 mm and flip-chip bonded onto a quartz substrate using epoxy (353ND, EPO-TEK). Then, the GaAs wafer is thinned to be ~20 μm using a mechanical lapping process and removed by a wet-etching process using citric acid etchant (citric acid : H2O2 = 5 : 1) until the wet etch stops at the Al0.55Ga0.45As stop-etch layer. The stop-etch layers are then wet-etched using a phosphoric acid etchant (H3PO4 : H2O : H2O2 = 20 : 200 : 4), and the GaAs spacer between them is etched using citric acid etchant. Finally, only the 400-nm-thick GaAs layer remains on the quartz substrate.

**Optical Measurements**

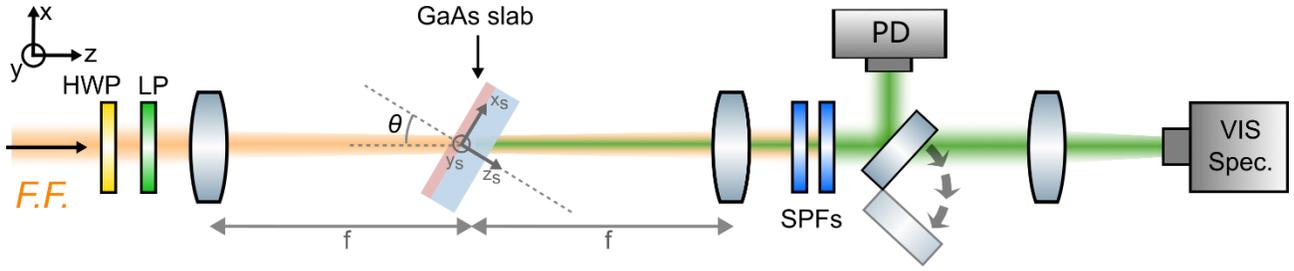

**Figure 8. Setup used for preliminary measurements shown in Fig.3.** See text for details.

The measurements shown in Fig. 3 were performed with a custom-built setup shown in Fig. 8. The pump signal at the FF, ranging between 1400 nm and 1750 nm, was a pulsed laser (pulse duration $\tau$ = 2 ps, repetition rate f = 80 MHz) provided by the signal of an optical parametric oscillator (APE, Levante IR ps) pumped by an Yb laser (APE, Emerald Engine). The power and polarization of the pump laser were controlled by a half-wave plate (HWP) and a linear polarizer (LP) cascaded in sequence. In all measurements in Fig. 3, the average power was kept fixed to about 400 mW (for all impinging angles and all wavelengths), corresponding to a pulse energy of about 5 nJ. The laser was then focused on the sample (from the quartz side) with a long-focal-length lens (focal length = 20 cm), resulting in a spot radius (defined as the distance from the center at which the intensity drops by a factor $e^2$) of approximately 90 $\mu m$. The signal emerging from the opposite side of the sample (containing both the FF and the SH signal) was collected by a second identical lens and re-collimated. After removing the FF with a pair of short-pass filters (SPF), the SH signal was redirected either to a visible spectrometer (Ocean Optics, HR4000) or to a photodiode (Thorlabs, DET100A2). The power of the laser (before exciting the sample) was continuously monitored by an additional germanium photodiode (not shown in Fig. 8).

To excite the sample with a pump with a well-defined *s* or *p*-polarization, we utilized the configuration shown in Fig. 8. The reference frame of the lab (labelled as *x,y,z*, black arrows in Fig. 8) is defined as such that *z* is along the optical axis of the setup, and *x* and *y* correspond to the *horizontal* and *vertical* polarizations, which can be imparted by LP. We also introduce the reference frame of the sample ($x_s, y_s, z_s,$ gray arrows), shown also in Fig. 3, defined such that $z_s$ is always perpendicular to the GaAs layer. The sample was mounted on two different rotation stages: a motorized one (Thorlabs, HDR50) to control the polar angle $\theta$, defined as the angle between the optical axis (*z*) and $z_s$ (see Fig. 8); and a manual rotation stage to control the azimuthal angle $\phi$, defined as the in-plane rotation of the sample around the axis $z_s$ (see also Fig. 3a). The rotational geometry (Fig. 8) is chosen such that, when $\theta = 0$ and $\phi = 0$, the reference frames of the lab (*x, y, z*) and of the sample ($x_s, y_s, z_s$) coincide (apart from a trivial translation). Moreover, for any value of $\theta$ and $\phi$, the vertical direction of the lab reference frame (*y*) is always parallel to the sample plane. Figure 8 shows the particular case of $\phi = 0$, for which the directions *y* and $y_s$ coincide. Thanks to this configuration, when the pump is polarized along the *y* direction of the lab frame (*vertical* polarization), the electric field of the pump always lies in the sample plane, for any value of $\theta$ and $\phi$. This corresponds to *s* polarization. Conversely, when the pump is polarized along the *x* direction of the lab frame (horizontal polarization), the magnetic field of the pump is along *y*, and thus it will always lie in the sample plane, for any value of $\theta$ and $\phi$. This corresponds to *p* polarization.

The measurements in Fig. 3(b-d) were performed by keeping the FF wavelength fixed at $\lambda_{FF} = 1550$ nm and by continuously varying the angles $\theta$ and $\phi$ of the sample while acquiring the SHG spectra with the spectrometer. The signal recorded by the spectrometer in a narrow region around $\lambda_{FF}/2$ was then used to retrieve the SHG efficiency. The procedure was repeated for two different orientations of the linear polarizer (LP), in order to set the pump polarization to either s or p as described above. The measurements in Fig. 3e were performed by varying the pump wavelength in the 1400 nm – 1750 nm range while keeping all other parameters fixed. The SHG signal produced by each pump was recorded by the photodiode, and the SHG efficiency (in arbitrary units) was obtained by dividing the voltage read by the photodiode by the square of the laser power (measured separately for each wavelength). The wavelength dependence of the photodiode efficiency (provided by the vendor) was used to correct the measured voltages.

The measurements in Figs. 5-7 were acquired with the setup shown in Fig. 4 and described in the main text. The laser source was the same as in the setup described in the previous paragraph.

**Funding.** Ministero dell'Istruzione e del Merito (METEOR, PRIN-2020 2020EY2LJT_002); H2020 Future and Emerging Technologies (FETOPEN-2018-2020 899673, METAFAST). Australian Research Council Centres of Excellence Program (CE200100010). I.B. and H.J. acknowledge support from the U.S. Department of Energy, Office of Basic Energy Sciences, Division of Materials Sciences and Engineering. This work was performed in part at the Center for Integrated Nanotechnologies, an Office of Science User Facility operated for the US Department of Energy (DOE) Office of Science. Sandia National Laboratories is a multimission laboratory managed and operated by National Technology and Engineering Solutions of Sandia, LLC., a wholly owned subsidiary of Honeywell International, Inc., for the U.S. Department of Energy's National Nuclear Security Administration under contract DE-NA0003525. This paper describes objective technical results and analysis. Any subjective views or opinions that might be expressed in the paper do not necessarily represent the views of the U.S. Department of Energy or the United States Government.

**Competing interests.** The authors declare no conflicts of interest.

**Contributions.** D.d.C., C.D.A., M.C., A.A. conceived the original idea and designed the experiment. D.d.C. and C.D.A. performed the theoretical analysis and simulations. H. J. and I. B. fabricated the sample.

M.C. built the optical setups, performed the measurements and analyzed the data. M.C., D.d.C., C.D.A., A.A. wrote the manuscript, with inputs from all authors.

**Corresponding author.** Correspondence to Andrea Alù.

**Data availability**. Data underlying the results presented in this paper may be obtained from the authors upon reasonable request.

| Linear Local | Linear Nonlocal | Nonlinear Nonlocal |
|---|---|---|
| $E_{out}^{\omega}(\mathbf{r}) = T_{LL}(\mathbf{r})E_{in}^{\omega}(\mathbf{r})$ | $E_{out}^{\omega}(\mathbf{k}) = T_{LN}(\mathbf{k})E_{in}^{\omega}(\mathbf{k})$ | $E_{out}^{2\omega}(\mathbf{k}) = \int d\mathbf{k}_1 T_{NN}(\mathbf{k}_1, \mathbf{k}-\mathbf{k}_1)E_{in}^{\omega}(\mathbf{k}_1)E_{in}^{\omega}(\mathbf{k}-\mathbf{k}_1)$ |
| | | $E_{in}^{\omega}(\mathbf{k}) = S^{\omega}(\mathbf{k}) + C^{\omega}(\mathbf{k})$ |

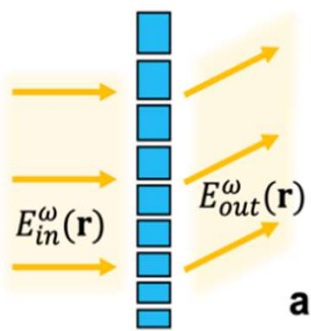
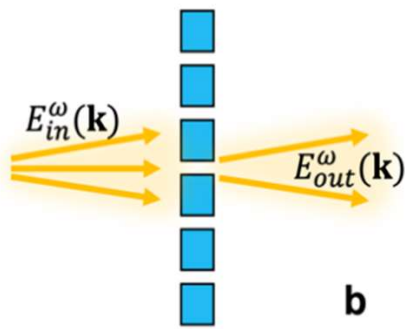
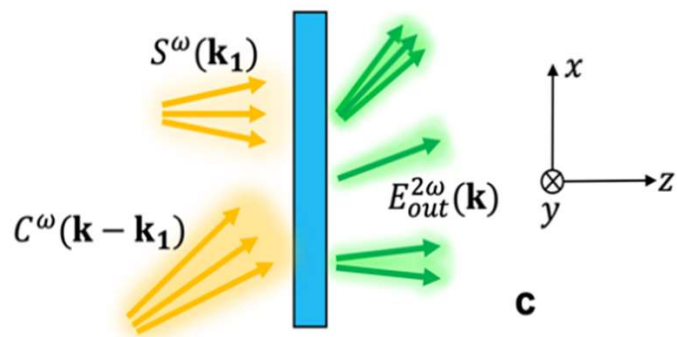

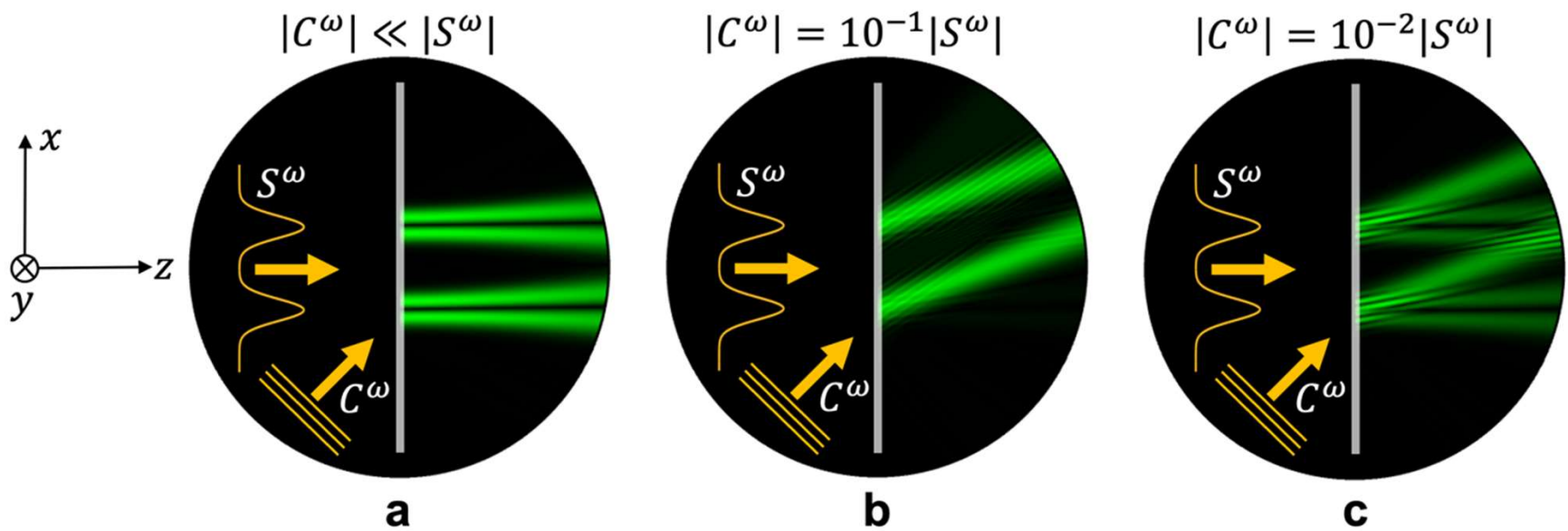

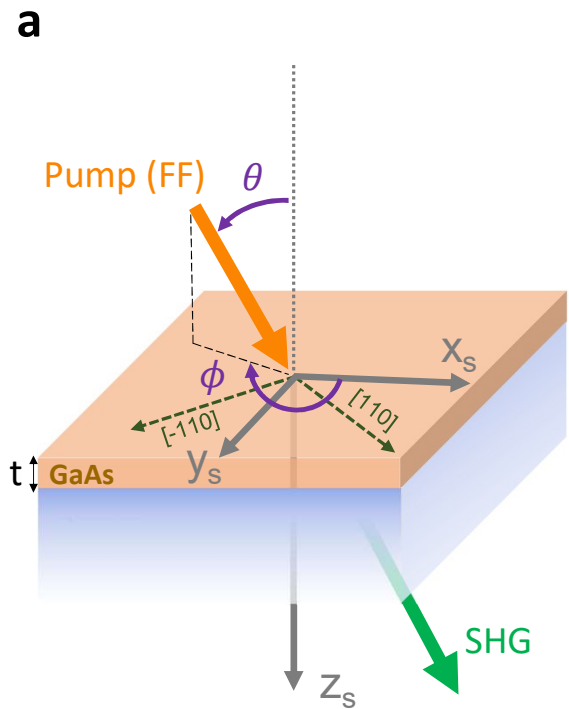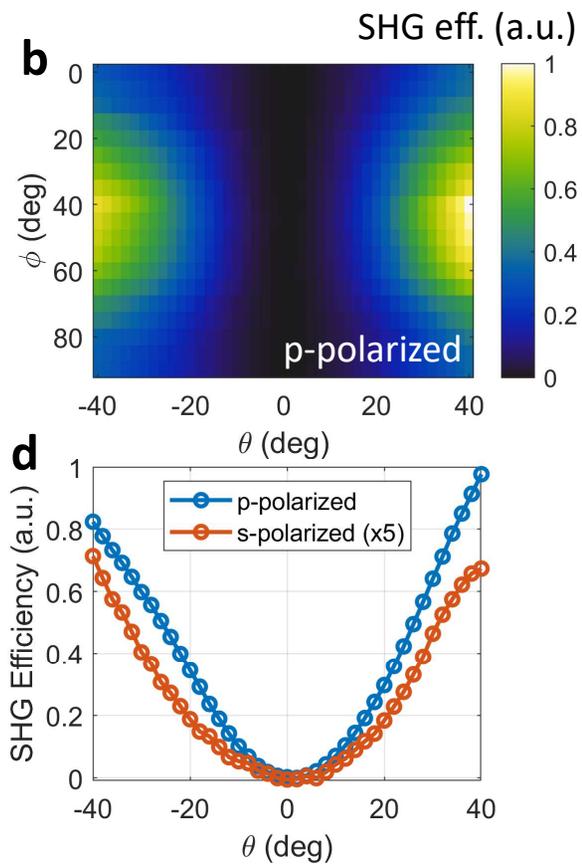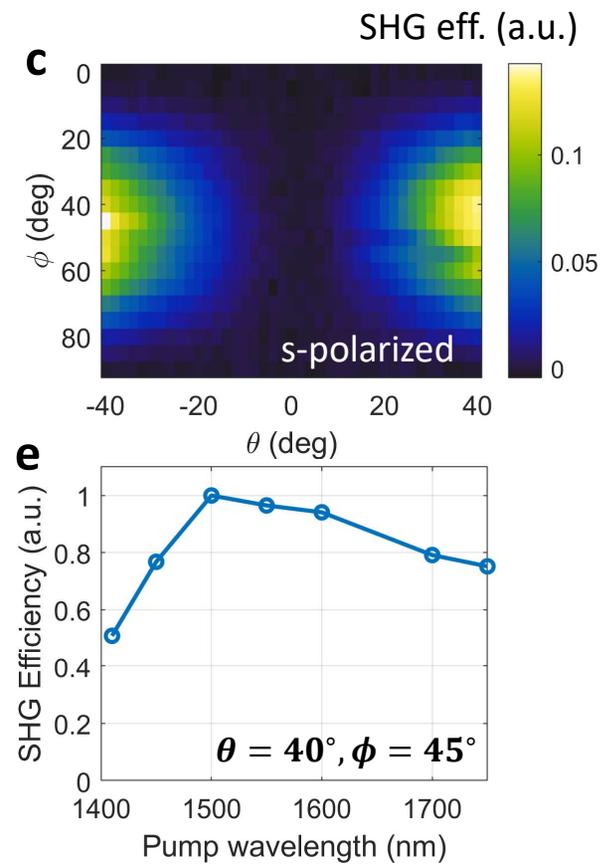

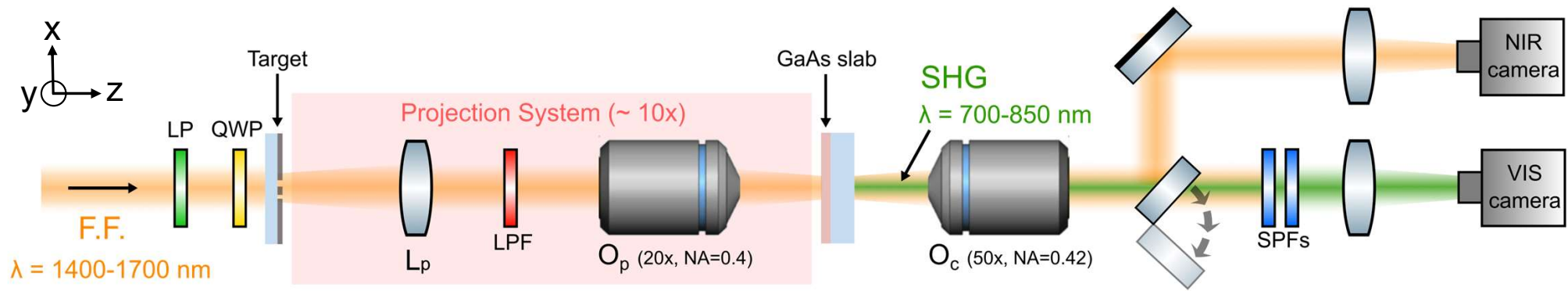

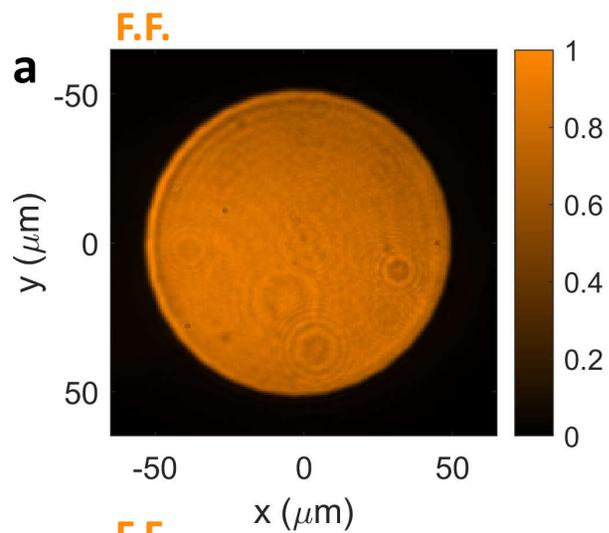 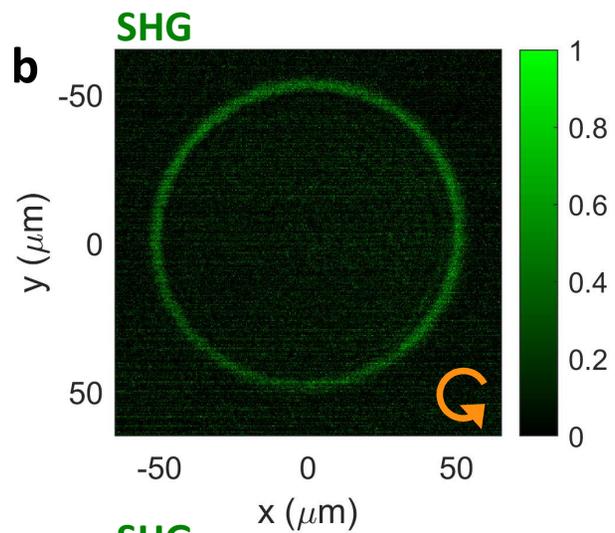
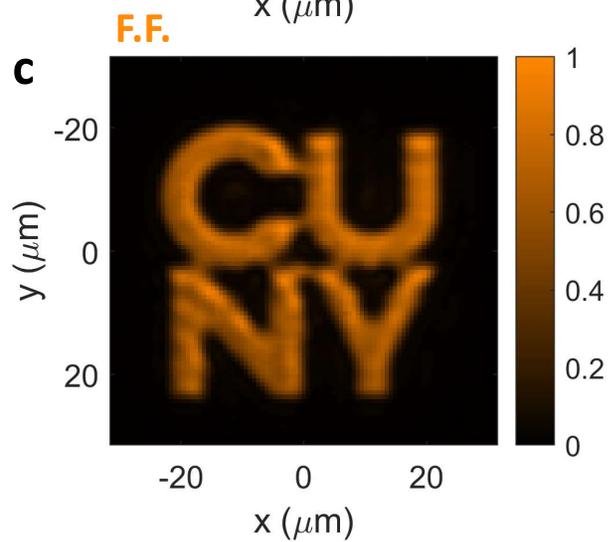 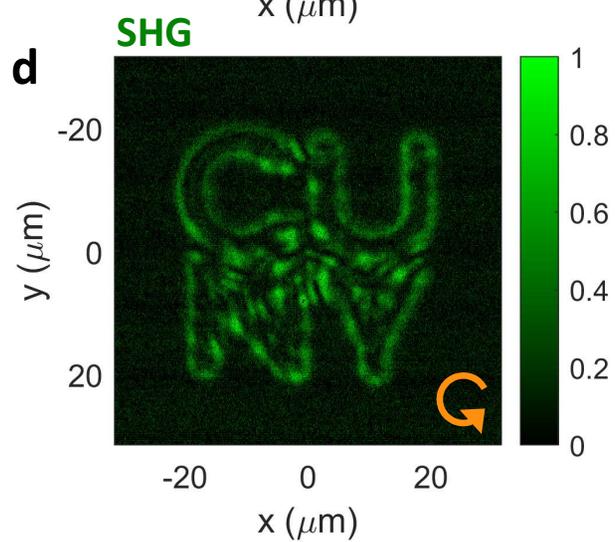

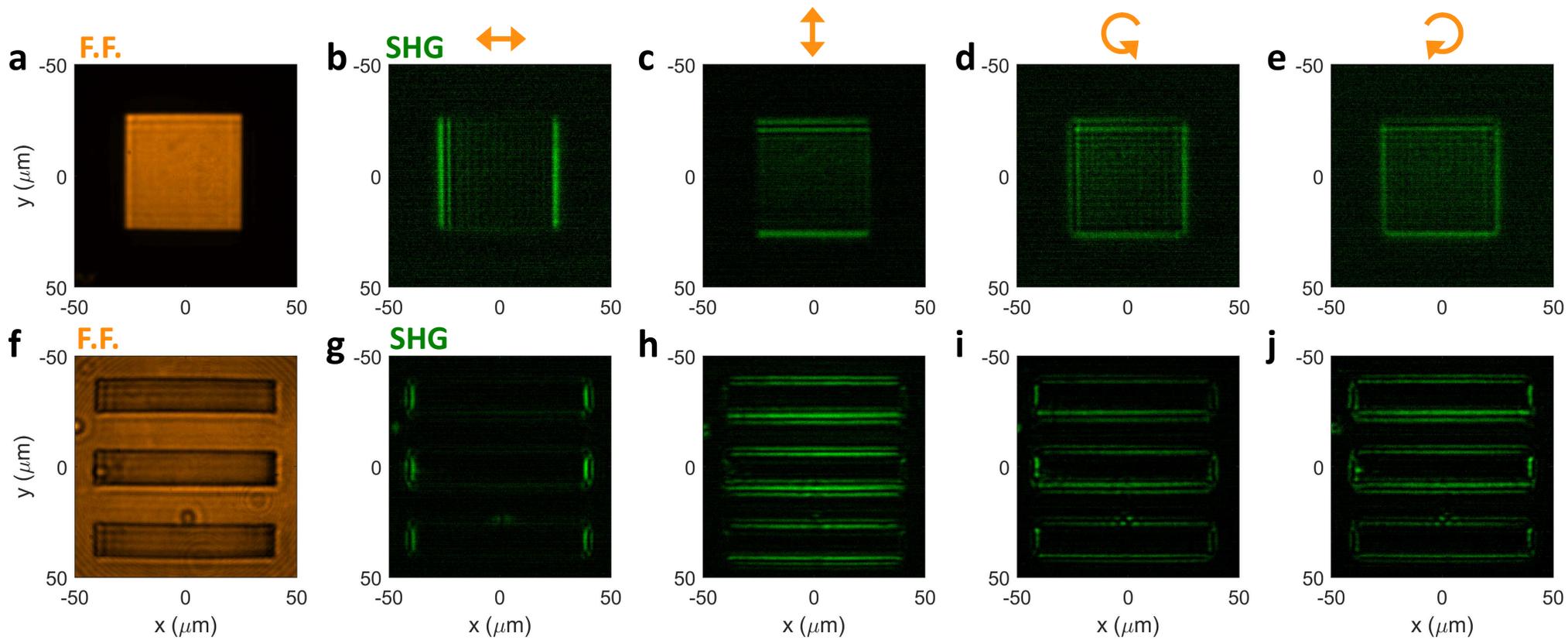

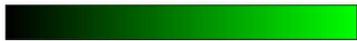
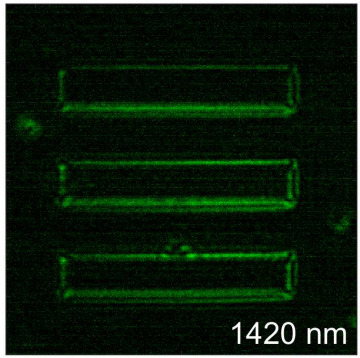
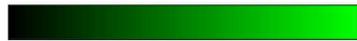
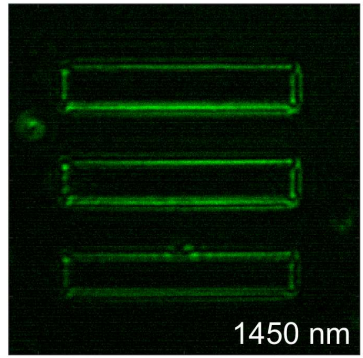
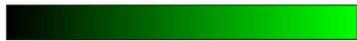
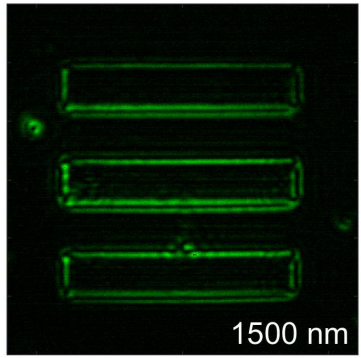
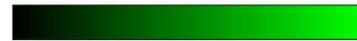
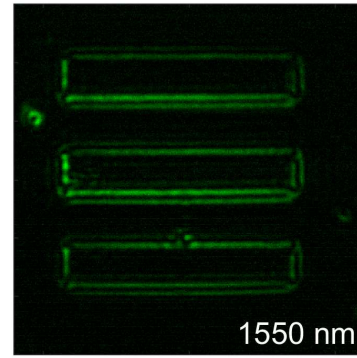
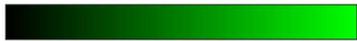
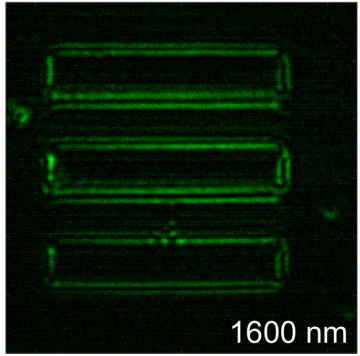
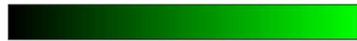
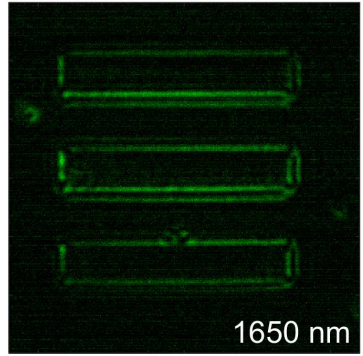
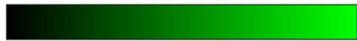
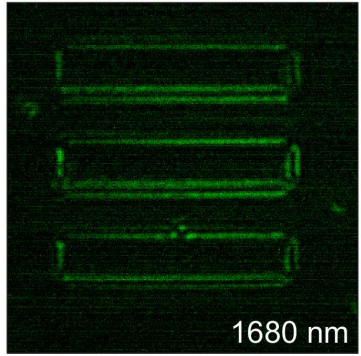

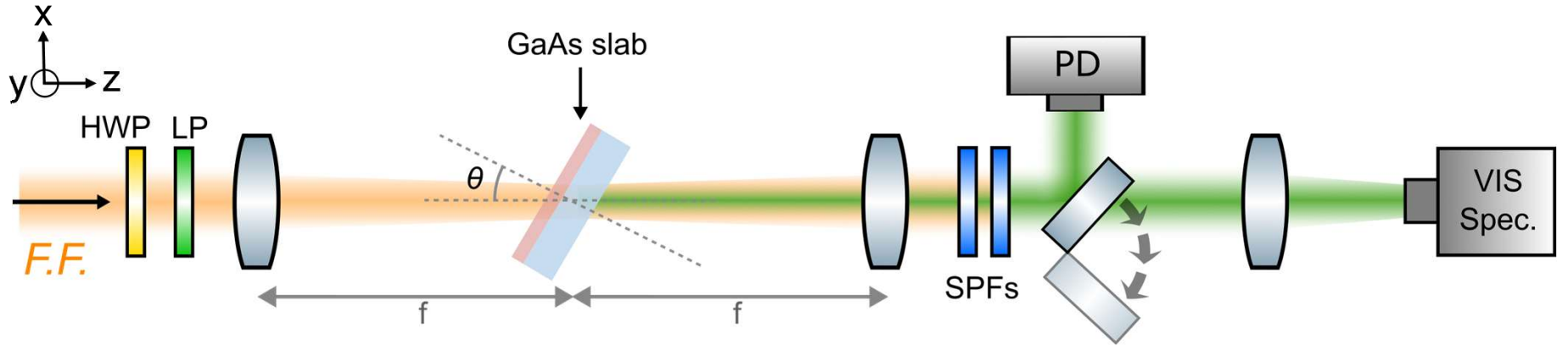

# Supplementary Information

## Table of Contents

## Section S1. Linear transmission spectrum of the sample – Calculations and Measurements

The finite thickness of the GaAs slab creates weak Fabry-Pérot resonances. As discussed in the paper and in the later sections of this SI, such resonances induce a weak spectral modulation of the SHG efficiency. To verify the magnitude of these effects, we have measured the normal-incidence transmission spectrum of the device. The measurement was performed by using a broadband supercontinuum laser (Leukos, Rock 400) filtered by a tunable narrowband filter (Photon, LLTF Contrast). The transmission spectrum was obtained by sweeping the laser wavelength and measuring the power transmitted through the metasurface with a power meter.

The measured transmission spectrum ((Fig. S1a, blue solid line) confirms the presence of strong oscillations, with a transmission contrast of $\Delta T > 40\%$ for wavelengths above 900 nm. For wavelengths below ~800 nm, the transmission suddenly drops, indicating the onset of strong absorption within the GaAs. We have numerically calculated the transmission spectrum of the GaAs slab with the transfer matrix method, by using tabulated values for the permittivity of GaAs (Palik) and assuming a slab thickness thickness of $t =$ 480 nm. The calculation (dashed red line in Fig. S1a) shows an excellent agreement with the measurements, correctly reproducing both the magnitude and period of the oscillations at long wavelengths and the onset of absorption at lower wavelengths. As expected, the transmission spectrum depends on the slab thickness (see calculations in Fig. S1b)

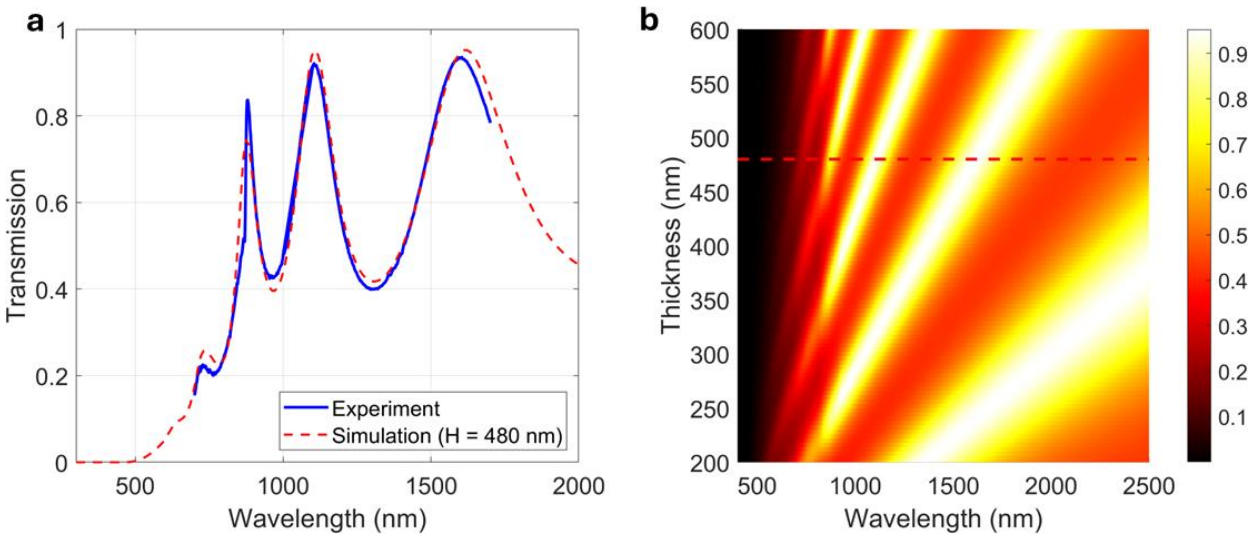

**Figure S1.** (a) Measured (blue solid line) and calculated (red dashed line) normal-incidence transmission spectrum of the device. In calculations, a GaAs slab thickness of $t =$ 480 nm was assumed. (b) Calculated normal-incidence transmission spectrum versus slab thickness. Red dashed line denotes the spectrum shown as red dashed line in panel a.

# Section S2. Numerical modeling of second harmonic generation from a GaAs film

The simulations of second-harmonic generation (shown in the main text and in the next sections of this SI) were performed in the undepleted pump approximation using the finite-element solver of COMSOL Multiphysics in the frequency domain. First, the pump field is calculated by solving the Helmholtz equation at the fundamental frequency using a plane-wave excitation with a given elevation (polar) angle θ and azimuthal angle ϕ. Next, the pump field is used to define the polarization density source inside the GaAs slab in the Helmholtz equation for the second-harmonic frequency, $P_i^{2\omega} = \epsilon_0 \chi_{ijk}^{(2)} E_j E_k$, where $i, j$ and $k$ are the crystal axes of GaAs, here assumed aligned with the laboratory axes, $\epsilon_0$ is the vacuum permittivity, $\chi_{ijk}^{(2)}$ are the entries of the nonlinear susceptibility tensor of GaAs. In crystals such as GaAs there is only one independent non-zero element of the susceptibility tensor, and the non-zero entries are those for which $i \neq j \neq k$. In other words, $\chi_{xyz}^{(2)} = \chi_{xzy}^{(2)} = \chi_{xyz}^{(2)} = \chi_{xyz}^{(2)} = \chi_{xyz}^{(2)} = \chi_{xyz}^{(2)} = \chi^{(2)}$, while all the other elements of the tensor are zero. The value of the $\chi^{(2)}$ depends on the pump wavelength. In particular, in our simulations we adopted the frequency-dependent data of $\chi^{(2)}$ reported in *S. Bergfeld, W. Daum, Phys. Rev. Lett. 90 (2003) 036801* and *M.L. Trolle, Theory of linear and nonlinear optical response: Zinc-blende semiconductors, Master's Thesis, Aalborg University, Denmark (2011)*. If the crystal axes are rotated around the [001] axis (z coordinate) by an azimuthal angle $\theta_c$ with respect to the laboratory axes, then the susceptibility is transformed by the rotation matrix $R_z(\theta_c)$, and it can be written in the laboratory frame as follows:

$$\bar{\chi}^{(2)}(\theta_c) = \chi^{(2)} \begin{pmatrix} \{0, 0, \sin 2\theta_c\} & \{0, 0, \cos 2\theta_c\} & \{\sin 2\theta_c, \cos 2\theta_c, 0\} \\ \{0, 0, \cos 2\theta_c\} & \{0, 0, -\sin 2\theta_c\} & \{\cos 2\theta_c, -\sin 2\theta_c, 0\} \\ \{\sin 2\theta_c, \cos 2\theta_c, 0\} & \{\cos 2\theta_c, -\sin 2\theta_c, 0\} & \{0, 0, 0\} \end{pmatrix}$$

For the particular case of $\theta_c = \pi/4$, corresponding to the experimental conditions in the imaging experiments reported in Figs. 5-7, the non-zero tensor elements are $\chi_{xxz}^{(2)}(\pi/4) = \chi_{xzx}^{(2)}(\pi/4) = -\chi_{yyz}^{(2)}(\pi/4) = -\chi_{yzy}^{(2)}(\pi/4) = \chi_{zxx}^{(2)}(\pi/4) = -\chi_{zyy}^{(2)}(\pi/4) = \chi^{(2)}$.

# Section S3. SHG efficiency versus pump polarization, pump direction and pump wavelength

In fig. 3 of the main paper, we show the measured SHG efficiency of the sample versus pump polarization, pump direction and pump wavelength. In this section, we complement the experimental results shown in the main text with numerical simulations. Figure S2(a-b) reproduces the experimental data shown in Figs. 3(b-c) of the main paper, i.e., the measured SHG efficiency for p-polarized pump (Fig. S2a) and s-polarized pump (Fig. S2b), as a function of the pump impinging direction. The impinging direction is identified here by the polar angle θ and the azimuthal angle ϕ, following the same reference frame as in Fig. 3 of the main paper. In Figs. S2c and S2d we show the corresponding simulated data, for the same polarization and range of impinging angles. An excellent agreement between the calculated and measured data is observed. For both measurements and calculations, we normalized the SHG efficiency to the maximum value, which occur for p-polarization at $\theta = 40°$.

Due to the cubic symmetry of GaAs, the SHG efficiency is periodic with respect to the angle ϕ with a period of 90°. This is displayed in Fig. S3a, which shows the same set of calculated data as in Fig. S2c but over a wider range of impinging angles. Fig. S2b shows a horizontal cross-section of Fig. S3a for $\phi = 45°$. The maximum peak for $\theta \sim 80°$ is associated with the minimum of pump reflection occurring at Brewster incidence for p-polarization.

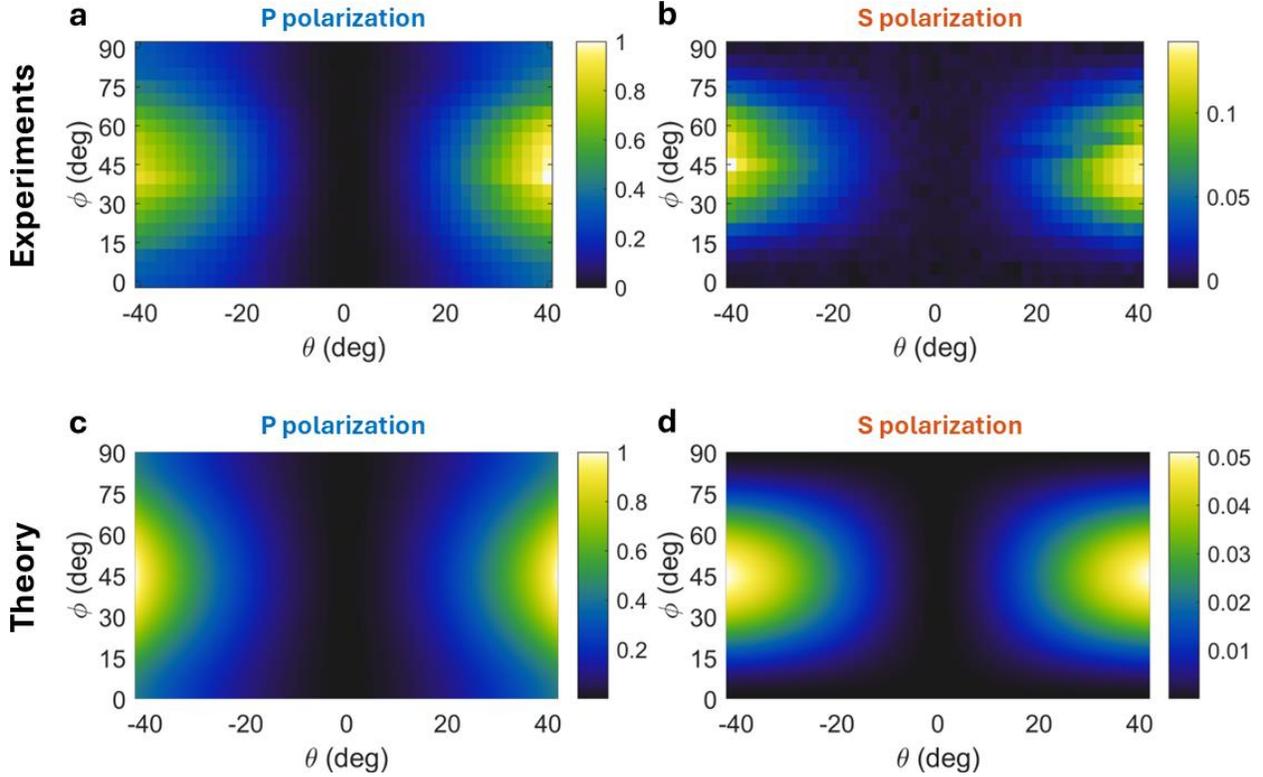

**Figure S2. Measured and Simulated SHG efficiency versus pump polarization and pump impinging direction.** The impinging direction is identified by the polar angle $\theta$ and the azimuthal angle $\phi$, following the same reference frame as in Fig. 3 of the main paper. **(a-b)** Measured SHG efficiency versus pump direction, for p polarization (panel a) and s polarization (panel b). **(c-d)** Calculated SHG efficiency versus pump direction, for p polarization (panel c) and s polarization (panel d).

Finally, Fig. S4a shows the calculated SHG efficiency versus pump wavelength, assuming a p-polarized pump impinging at $\theta = 40°, \phi = 45°$. As expected, the SHG efficiency peaks in the NIR region, and it is modulated by both the Fabry-Pérot resonances described above and by the intrinsic spectral dispersion of the $\chi^{(2)}$ of GaAs. For pump wavelengths below 1400 nm (corresponding to SH wavelengths below 700 nm) the SHG efficiency drops substantially due to the increased absorption of the SH inside the GaAs slab. Figure S4b shows a zoomed-in view of panel a in the range [1300 nm – 1800 nm], together with the experimentally measured efficiency (orange circles) that was reported in Fig. 3e of the main paper. Both measurements and calculations confirm the presence of a local maximum in the [1500 nm – 1600 nm] region, and a smooth decrease of the efficiency for wavelengths below or above this range.

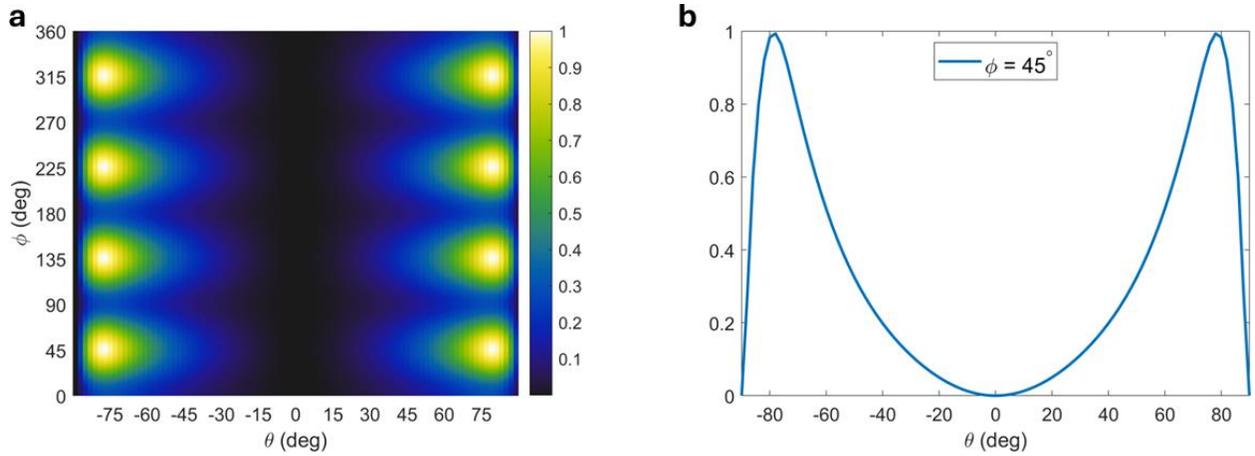

**Figure S3. Calculated SHG efficiency for p-polarized pump. (a)** Same data as in Fig. S2c but shown on a wider range of values of $\theta$ and $\phi$. **(b)** Horizontal cross-section of panel a, for $\phi = 45°$.

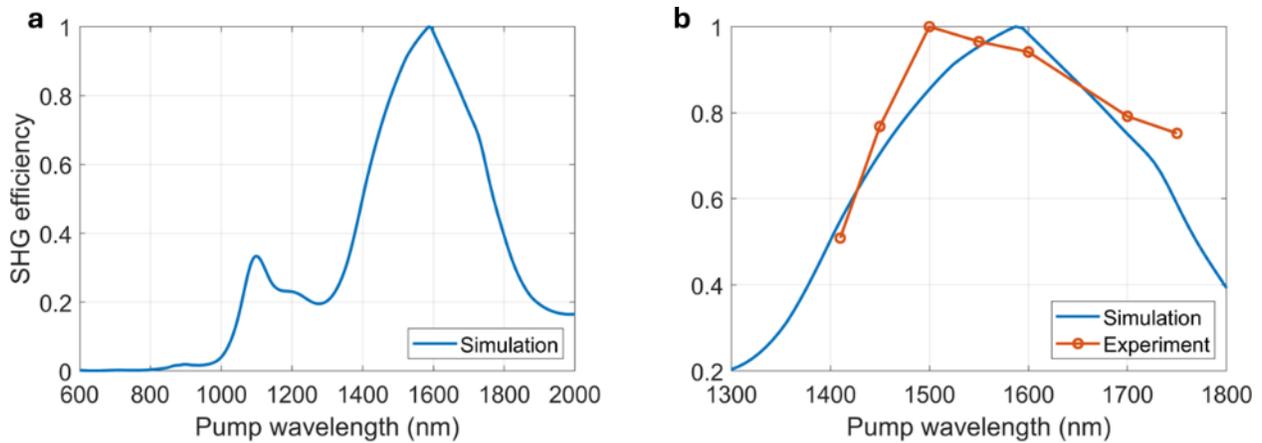

**Figure S4. SHG efficiency versus wavelength. (a)** Calculated SHG efficiency (in arbitrary units) for a p-polarized pump impinging at $\theta = 40°, \phi = 45°$, versus the pump wavelength. **(b)** Zoomed-in view of panel a in the [1300 nm – 1800 nm] range. The orange line corresponds to the experimentally measured efficiency reported in Fig. 3e of the main paper. Both measured and calculated efficiencies have been normalized such that their maximum is equal to 1.

## Section S4. Nonlinear edge detection under different crystal orientations

Here we show how the edge detection is affected by the orientation of the GaAs crystal with respect to the basis of liner polarization of the input image. In particular, we consider edge detection under two different scenarios: (i) the crystal axes are aligned to the laboratory axes; (ii) the *xy* crystal axes are rotated by 45° with respect to the *xy* laboratory axes. For both scenarios, the edge detection is evaluated under linear and circular polarization pumps, defined with respect to the laboratory frame. As a benchmark input object, an input aperture with octagonal shape is chosen, as illustrated in Fig. S5. To better highlight the edges detected at the second harmonic from the GaAs film, the octagon has smooth edges, obtained with a gaussian low-pass filter.

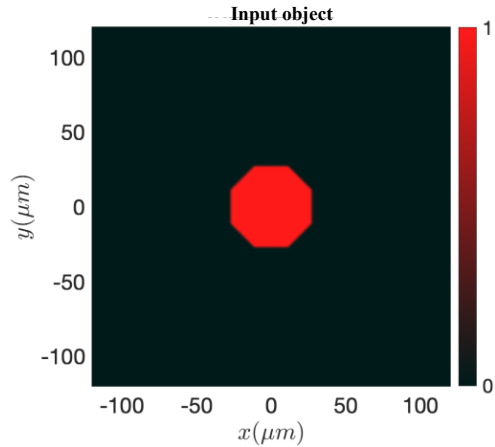

**Figure S5. Input object with octagonal shape for nonlinear edge detection.** An aperture with octagonal shape is used to test nonlinear edge detection under different crystal axes orientations and different pump polarization states.

In Fig. S6 we show the SHG image emerging from the GaAs film in the first scenario (crystal and laboratory frames aligned). While for circularly polarized pumps all the edges of the octagon are enhanced with equal intensity, the edge detection for linearly polarized pumps shows a pronounced sensitivity to the polarization orientation, in accordance with the experimental data shown in Fig. 6 of the main text. In particular, the edges detected with largest intensity, i.e., with maximum second-harmonic generation efficiency, are those perpendicular to the pump polarization orientation. Edges parallel to the pump polarization are weakly detected. In the particular case in which the polarization axis of the pump is aligned to one of the crystal axes, the second-harmonic intensity of the edges parallel to the pump polarization vanish. The same behavior, with edge detection sensitivity rotated by 45°, can be observed in the second scenario (illustrated in Fig. S7), in which the crystal axes rotated by 45° with respect to the laboratory axes.

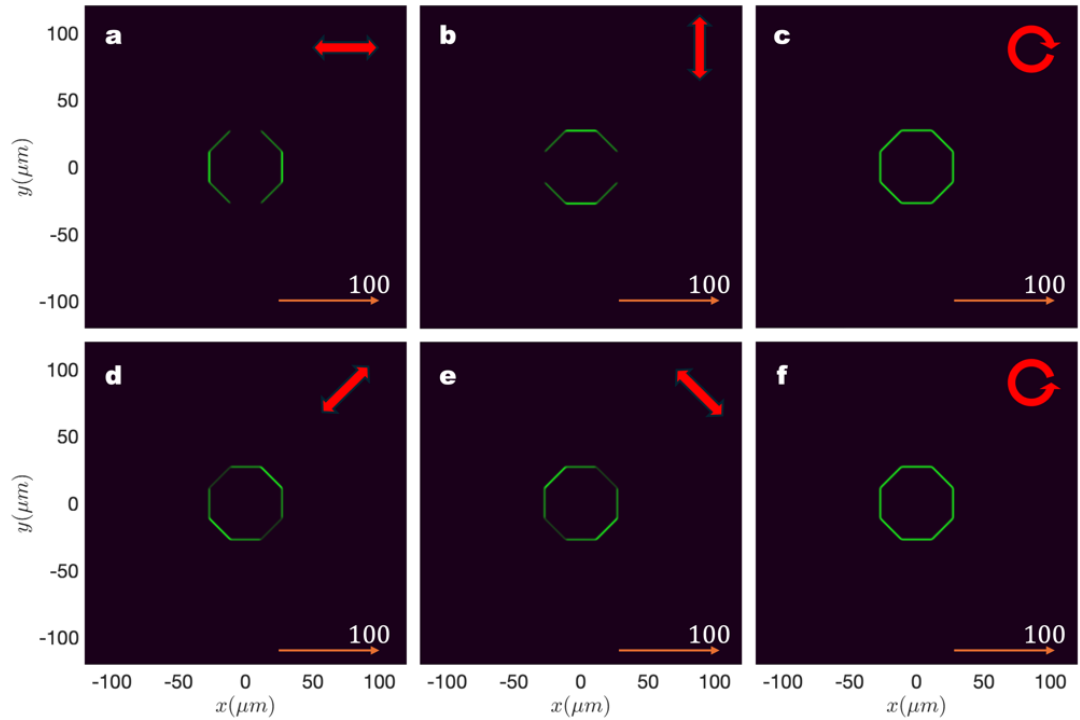

**Figure S6. Nonlinear edge detection of an octagonal shape when the crystal axes are aligned to the laboratory axes.** **(a)** Pump horizontally polarized ($\hat{x}$) and parallel to the [100] axis of the crystal. Edges along this direction are not detected. **(b)** Pump vertically polarized ($\hat{y}$) and perpendicular to the [100] axis of the crystal. Edges along this direction are not detected. **(c)** Pump with right circular polarization [$(\hat{x} + i\hat{y})/\sqrt{2}$]. **(d)** Pump linearly polarized at 45° with respect to the [100] crystal axis [$(\hat{x} + \hat{y})/\sqrt{2}$]. Edges along this direction are weakly detected. **(e)** Pump linearly polarized at 45° with respect to the [100] crystal axis [$(\hat{x} + \hat{y})/\sqrt{2}$]. Edges along this direction are weakly detected. **(f)** Pump linearly polarized at -45° with respect to the [100] crystal axis [$(\hat{x} - \hat{y})/\sqrt{2}$]. Edges along this direction are weakly detected **(f)** Pump with left circular polarization [$(\hat{x} - i\hat{y})/\sqrt{2}$].

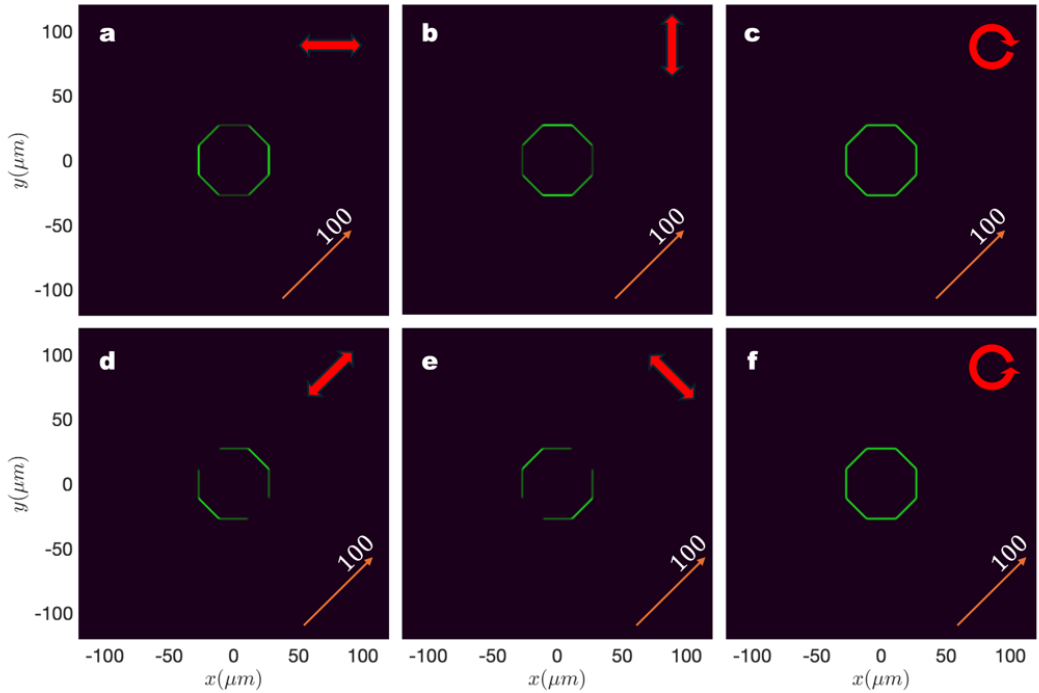

**Figure S7. Nonlinear edge detection of an octagonal shape when the crystal axes are rotated by 45° with respect to the laboratory axes. (a)** Pump horizontally polarized ($\hat{x}$), at -45° with respect to the [100] crystal axis. Edges along this direction are weakly detected. **(b)** Pump vertically polarized ($\hat{y}$) at +45° with respect to the [100] crystal axis. Edges along this direction are weakly detected. **(c)** Pump with right circular polarization $[(\hat{x} + i\hat{y})/\sqrt{2}]$. **(d)** Pump linearly polarized at 45°, parallel to the [100] crystal axis $[(\hat{x} + \hat{y})/\sqrt{2}]$. Edges along this direction are not detected. **(e)** Pump linearly polarized at -45°, perpendicular to the [100] crystal axis $[(\hat{x} - \hat{y})/\sqrt{2}]$. Edges along this direction are not detected. **(f)** Pump with left circular polarization $[(\hat{x} - i\hat{y})/\sqrt{2}]$.